\documentclass[usenatbib]{mnras}
\usepackage{newtxtext,newtxmath}
\usepackage[T1]{fontenc}
\usepackage{ae,aecompl}

\usepackage{graphicx}	% Including figure files
\usepackage{amsmath}	% Advanced maths commands
\usepackage{subcaption}
\graphicspath{{./figs/}}
\DeclareGraphicsExtensions{.pdf,.png}

\newcommand{\kpc}{\,\textrm{kpc}} % math mode kpc
\newcommand{\myr}{\,\textrm{Myr}} % math mode Myr
\newcommand{\rcore}{r_\textrm{c}} % math mode Myr
\newcommand{\uvec}[1]{\boldsymbol{\hat{\textbf{#1}}}}

\renewcommand{\vec}[1]{\boldsymbol{#1}}

\title[Jets in asymmetric environments]{Dynamics of relativistic radio jets in asymmetric environments}

\author[Patrick M. Yates-Jones et al.]{
Patrick M. Yates-Jones,$^{1,2,3}$\thanks{E-mail: patrick.yates@utas.edu.au}
Stanislav S. Shabala,$^{1,2,3}$
Martin G. H. Krause$^{2,1}$
\\
$^{1}$ School of Natural Sciences, Private Bag 37, University of Tasmania, Hobart, TAS 7001, Australia\\
$^{2}$ Centre for Astrophysics Research, University of Hertfordshire, College Lane, Hatfield, Herts AL10 9AB, UK \\
$^{3}$ ARC Centre of Excellence for All Sky Astrophysics in 3 Dimensions (ASTRO 3D) \\
}

\date{Accepted XXX. Received YYY; in original form ZZZ}

\pubyear{2021}

\begin{document}
\label{firstpage}
\pagerange{\pageref{firstpage}--\pageref{lastpage}}
\maketitle

\begin{abstract}
  We have carried out relativistic three-dimensional simulations of high-power radio sources propagating into asymmetric cluster environments. 
  We offset the environment by $0$ or $1$ core radii (equal to $144\kpc$), and incline the jets by $0$, $15$, or $45\degr$ away from the environment centre.
  The different environment encountered by each radio lobe provides a unique opportunity to study the effect of environment on otherwise identical jets.
  We find that the jets become unstable towards the end of the simulations, even with a Lorentz factor of $5$; they nevertheless develop typical FR II radio morphology.
  The jets propagating into denser environments have consistently shorter lobe lengths and brighter hotspots, while the axial ratio of the two lobes is similar.
  We reproduce the recently reported observational anti-correlation between lobe length asymmetry and environment asymmetry, corroborating the notion that observed large-scale radio lobe asymmetry can be driven by differences in the underlying environment.
\end{abstract}


%%%%%%%%%%%%%%%%%%%%%%%%%%%%%%%%%%%%%%%%%%%%%%%%%%

%%%%%%%%%%%%%%%%% BODY OF PAPER %%%%%%%%%%%%%%%%%%

\section{Introduction}\label{sec:intro}

  Radio sources are inextricably linked to their host galaxy through feedback processes \citep{McNamara2007,Fabian2012,Hardcastle2020} which influence black hole growth \citep{Rafferty2006,Dubois2012,Alexander2012}, star formation \citep{Gaibler2012,Weinberger2017}, and the small to large-scale environment through gas circulation, heating, and shock driving \citep{Croton2006,Rafferty2008,Mukherjee2016,English2016,Yates2018}.
  Their influence on galaxy evolution and formation is underscored by the significant attention paid towards analytic \citep{Scheuer1974,Begelman1989,Falle1991,Kaiser1997,Alexander2002,Alexander2006a,Shabala2009,Krause2012,Raouf2017}, semi-analytic \citep{Turner2015,Hardcastle2018,Raouf2019}, and numerical \citep{Krause2005,Gaibler2011a,Wagner2012a,Hardcastle2014,Mukherjee2016,Bicknell2018,Yates2018,Li2018a,Perucho2019b,Horton2020} models to understand this phenomenon.
  These radio sources are most commonly studied through observations of the kpc to Mpc scale radio lobes \citep{Bennett1962,Hardcastle2019,Seymour2020} inflated by the relativistic jet beam, throughout which a population of relativistic electrons accelerated at shocks emits through synchrotron and inverse-Compton processes \citep{Kaiser1997a,Turner2018a}

  Radio sources have historically been classified into two distinct categories \citep{Fanaroff1974} based on whether the large-scale lobe morphology is edge-darkened (Fanaroff-Riley class I, FR I) or edge-brightened (FR II).
  As radio telescope sensitivity and resolution have increased, this distinction has become increasingly murky: compact radio sources with some low-frequency extended emission have been termed FR 0s \citep{Garofalo2019,Baldi2019a,Capetti2019}; low luminosity ($L_{150} \le 10^{25}\,\textrm{W Hz}^{-1}$) FR II radio sources have been observed \citep{Mingo2019}; radio sources which exhibit hybrid FR I/II characteristics suggest that precession and projection effects can drastically alter the observed morphology \citep{Harwood2020,Krause2019a,Horton2020a,Horton2020}; the consideration of mergers, cluster weather, and remnant or restarting radio sources further complicates the observed lobe morphology \citep{Mahatma2018,Yates2018,English2019,Joshi2019,Hardcastle2019a,ONeill2019,Bruni2020}.

  Many complex processes are likely at work behind the observed morphologies, however, one key factor is the environment into which the radio source is expanding.
  Analytic models predict a strong dependence on lobe size as a function of environment \citep{Kaiser1997,Alexander2006a,Krause2012}, which is reproduced by simulations \citep{Mendygral2012,Hardcastle2013,Bourne2017,Yates2018}.
  Observationally, the observed length asymmetry in radio lobes pairs was recently shown to be linked to the underlying environment asymmetry, as traced by optical galaxy counts by \citet{Rodman2019}, who found a statistically significant anti-correlation between the two.

  At first sight, it might sound trivial that jets propagate more slowly into a higher density environment.
  However, the environment sets the pressure level in the radio lobes, which in turn affects the hydrodynamic collimation of the jet \citep[e.g.,][]{Kaiser1997}.
  Jets in a denser environment have higher pressure lobes and narrower jets, which concentrate the thrust in a smaller area of the jet head and thus make it propagate faster and change the area over which feedback is distributed.
  While the sound speed in radio lobes is high such that, to first order, pressure differences are quickly readjusting, it is well known that dynamically changing as well as quasi-persistent pressure gradients exist, for example away from hotspots \citep[Fig. 12]{Krause2003}.
  There is thus a non-linear feedback loop, and only a simulation that includes all these relevant processes can make a proper prediction for this situation.

  In this work, we confront the findings of \citet{Rodman2019} with hydrodynamic simulations of radio jets in asymmetric environments and show that the observed radio source asymmetry is consistent with being caused by the underlying environment asymmetry.
  \autoref{sec:setup} describes the technical simulation setup, environment prescription, and parameter space explored.
  In \autoref{sec:results} we present our findings, and our conclusions in \autoref{sec:conclusions}.

  Throughout this work, we adopt the Planck15 cosmology \citep{Planck2016} with $H_0 = 67.6\,\textrm{km s}^{-1}\,\textrm{Mpc}^{-1}$ and $\Omega_\textrm{M} = 0.307$.

\section{Simulations}\label{sec:setup}

  \begin{figure*}
    \includegraphics{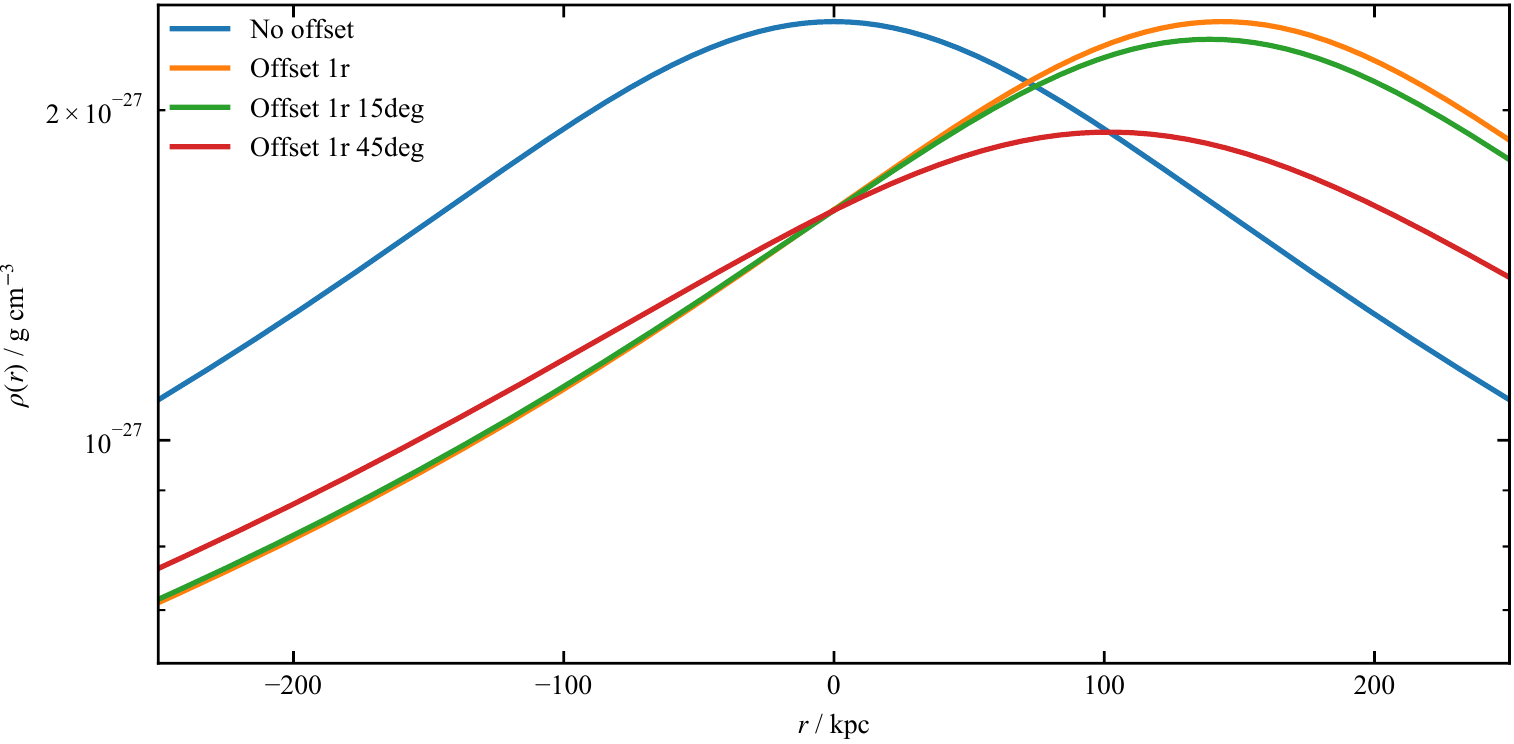}
    \caption{
      Environment density as a function of radius along the jet axis for the environment parameters described in \autoref{sec:environment}.
      The environment centre is offset by $0$ or $1$ core radii, while the jet inclination angle with respect to the environment is $0$, $15$, or $45\degr$.
    }
    \label{fig:environment_density}
  \end{figure*}

  \begin{figure}
    \def\svgwidth{\columnwidth}
    %% Creator: Inkscape 1.0.2 (e86c870879, 2021-01-15, custom), www.inkscape.org
%% Accompanies image file 'injection-zone-spherical.pdf' (pdf, eps, ps)
\begingroup%
  \makeatletter%
  \providecommand\color[2][]{%
    \errmessage{(Inkscape) Color is used for the text in Inkscape, but the package 'color.sty' is not loaded}%
    \renewcommand\color[2][]{}%
  }%
  \providecommand\transparent[1]{%
    \errmessage{(Inkscape) Transparency is used (non-zero) for the text in Inkscape, but the package 'transparent.sty' is not loaded}%
    \renewcommand\transparent[1]{}%
  }%
  \providecommand\rotatebox[2]{#2}%
  \newcommand*\fsize{\dimexpr\f@size pt\relax}%
  \newcommand*\lineheight[1]{\fontsize{\fsize}{#1\fsize}\selectfont}%
  \ifx\svgwidth\undefined%
    \setlength{\unitlength}{239.91126385bp}%
    \ifx\svgscale\undefined%
      \relax%
    \else%
      \setlength{\unitlength}{\unitlength * \real{\svgscale}}%
    \fi%
  \else%
    \setlength{\unitlength}{\svgwidth}%
  \fi%
  \global\let\svgwidth\undefined%
  \global\let\svgscale\undefined%
  \makeatother%
  \begin{picture}(1,1.27457792)%
    \lineheight{1}%
    \setlength\tabcolsep{0pt}%
    \put(0,0){\includegraphics[width=\unitlength,page=1]{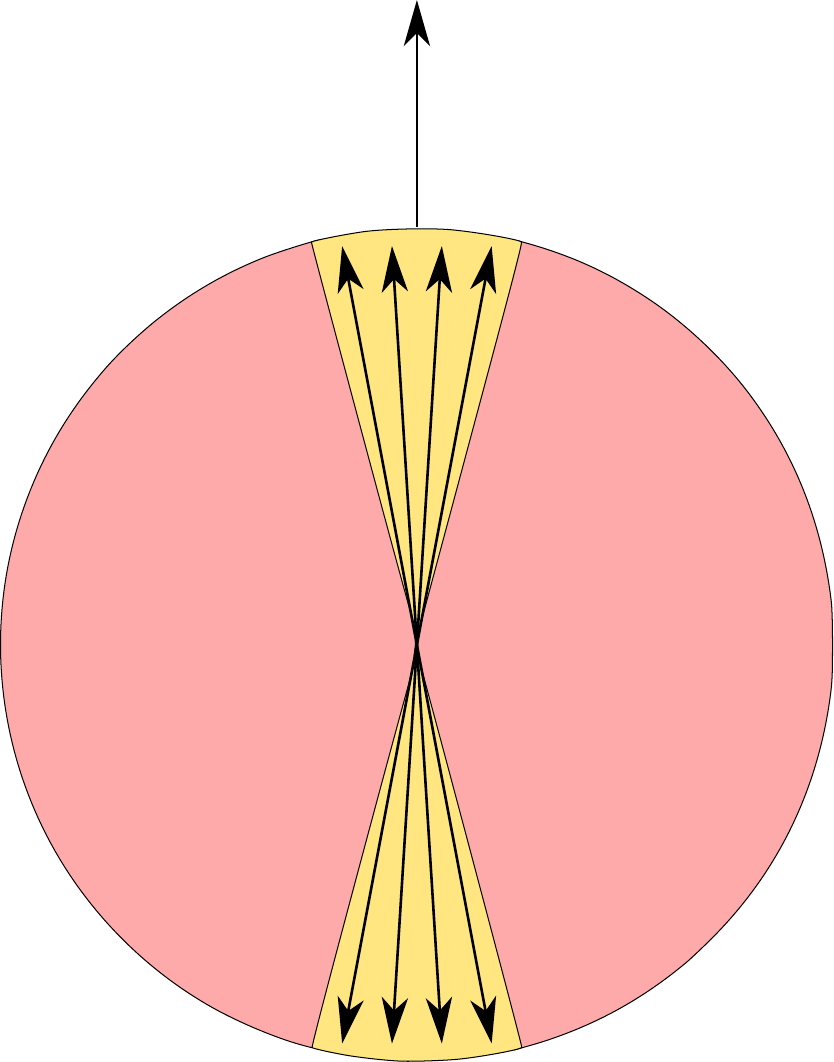}}%
    \put(0.1769392,0.71679465){\makebox(0,0)[lt]{\lineheight{1.25}\smash{\begin{tabular}[t]{l}$\rho(r)$\\$P(r)$\end{tabular}}}}%
    \put(0.53755584,1.02867317){\makebox(0,0)[lt]{\lineheight{1.25}\smash{\begin{tabular}[t]{l}$v_r$\end{tabular}}}}%
    \put(0.50375101,0.4866419){\makebox(0,0)[lt]{\lineheight{1.25}\smash{\begin{tabular}[t]{l}$(0, 0, 0)$\end{tabular}}}}%
    \put(0,0){\includegraphics[width=\unitlength,page=2]{injection-zone-spherical.pdf}}%
    \put(0.21542171,0.51938032){\makebox(0,0)[lt]{\lineheight{1.25}\smash{\begin{tabular}[t]{l}$r_0$\end{tabular}}}}%
    \put(0.53853967,1.22682633){\makebox(0,0)[lt]{\lineheight{1.25}\smash{\begin{tabular}[t]{l}$Z$\end{tabular}}}}%
  \end{picture}%
\endgroup%

    \caption{
      Jet injection zone.
    }
    \label{fig:injection_zone}
  \end{figure}

  The simulations presented in this paper model astrophysical jets as relativistic fluid flows using the freely available numerical simulation package \textsc{pluto}\footnote{\url{http://plutocode.ph.unito.it/}} \citep{Mignone2007}.
  We use a simulation setup that builds upon our earlier work \citep{Yates2018}, extended to model relativistic jets in three dimensions.

  \subsection{Environment}\label{sec:environment}

    A key feature of this modelling is the observationally motivated environments used as the initial conditions for the simulations.
    Our focus is on low-redshift clusters, which are reasonably well described by the isothermal beta profile \citep{Cavaliere1978} for gas density as a function of radius from the core $\rho(r)$, where
    \begin{equation}
      \rho(r) = \rho_0 \left[ 1 + \left( \frac{r}{r_\textrm{c}} \right)^2 \right] ^ {-3 \beta / 2}\,.
    \end{equation}\label{eqn:density-isothermal-beta}
    This profile is parameterised by the core density $\rho_0$, core radius $r_\textrm{c}$, and power-law slope $\beta$.
    In this work, we use a density profile typical of a cluster environment, with $M_\textrm{halo} = 3\times10^{14}\,\textrm{M}_{\sun}$ and $\beta = 0.38$.
    We set the core radius to be $0.1 R_\textrm{vir}$, or $r_c = 144\,\textrm{kpc}$, following \citet{Yates2018}; observationally the core radii of clusters vary between tens and hundreds of kpc \citep{Vikhlinin2006}.
    These parameters give a central density of $\rho_0 = 1.4\times 10^{-27}\,\textrm{g cm}^{-3}$.
    In \autoref{fig:environment_density} we show the environment density profile as a function of distance along the jet axis, for the different offset and inclination angles used in this work.
    Jets launched offset from the cluster centre experience significant asymmetry in the environment density profile; in \autoref{sec:dynamics_and_morphology} we argue that this asymmetry has a large impact on the resulting jet dynamics.
    We also note that the impact of inclination angle on density is less pronounced, however, an inclined jet does experience a gravitational acceleration asymmetry which affects remnant morphology.

    The pressure profile is calculated from the density profile using $P(r) = \frac{k_\textrm{B} T \rho(r)}{\mu m_\textrm{H}}$.
    We take $\mu = 0.60$ and $T = 3.46\times 10^7\,\textrm{K}$, consistent with observed temperatures in clusters.
    Cooling is not included in our simulations.
    This is a valid assumption provided the environment cooling time is long; the central cooling time $t_\textrm{cool}$ for the environment presented here is on the order of $1.5\,\textrm{Gyr}$ for a metallicity of $Z = -1.0$ \citep{Sutherland1993}, more than an order of magnitude longer than the jet lifetimes considered in this work.

    The radial gravitational acceleration necessary to keep the environment in hydrostatic equilibrium is
    \begin{equation}
      \vec{g} = - 3.0\, \beta \frac{c_\textrm{s}^2}{\gamma} \frac{r}{r_\textrm{c}^2 + r^2} \uvec{r}\,,
      \label{eqn:gravity_isothermal_beta}
    \end{equation}
    where $c_\textrm{s} = \left(\frac{ \Gamma P }{ \rho }\right)^{1/2}$ is the environment sound speed.
    This is implemented as a source term in \textsc{pluto}.
    The stability of the environment is confirmed by evolving it for over $100\,\textrm{Myr}$, twice as long as the maximum jet time.

    We offset the environment in both the Y and Z directions to study the parameter space detailed in \autoref{sec:parameter_study}.
    This is achieved through a transformation of coordinates with respect to the new environment centre, resulting in an offset radius given by
    \begin{equation}
      r' = \sqrt{(x - x_\textrm{offset})^2 + (y - y_\textrm{offset})^2 + (z - z_\textrm{offset})^2} \label{eqn:offset_radius}\,.
    \end{equation} 

  \subsection{Computational methods}\label{sec:technical_details} 

    Version 4.3 of \textsc{pluto} is used to carry out the simulations presented here.
    The relativistic hydrodynamics physics module is used, along with the \textsc{hllc} Riemann solver, linear reconstruction, second-order Runge-Kutta time-stepping, the Taub-Mathews equation of state, and a Courant-Friedrichs-Lewy (CFL) number of $0.33$.

    At each timestep \textsc{pluto} integrates the system of conservation laws described as
    \begin{align}
      \frac{\upartial}{\upartial t}
        \begin{pmatrix}
          D \\ \vec{m} \\ \vec{E}
        \end{pmatrix}
        + \nabla \cdot
        \begin{pmatrix}
          D \vec{v} \\ \vec{m} \vec{v} + p \mathbf{I}\\ \vec{m}
        \end{pmatrix}
        = 0\,,
    \end{align}
    with conservative state variables $D, \vec{m}, E$ (the laboratory density, momentum density, and total energy density respectively), fluid three-velocity $\vec{v}$, and pressure $p$.

    The simulations were carried out on a static three-dimensional Cartesian grid centred at the origin with a side length of $l_\textrm{x,y,z} = 400\,\textrm{kpc}$, cell count of $(n_\textrm{x}, n_\textrm{y}, n_\textrm{z}) = (550, 550, 960)$, and reflective boundary conditions.
    A central uniform grid patch of $100$ cells is defined around the injection region in all three dimensions ($-2.5 \rightarrow 2.5\,\textrm{kpc}$, a resolution of $0.05\,\textrm{kpc}$) to ensure that injection of the initially conical jet and its hydrodynamic collimation by ambient pressure are sufficiently resolved.
    Either side of the central grid patch is a geometrically stretched grid consisting of $430$ cells in Z, and $225$ cells in X and Y; giving a minimum resolution of $1.7\kpc$ and $2.8\kpc$ per cell respectively.
    Reflective boundary conditions were applied to both the lower and upper grid boundaries.

    A two-dimensional spherical axisymmetric grid is used in a resolution study to verify that jet recollimation is captured correctly.
    In these simulations the jet is aligned with the axis of symmetry, and the simulation domain is taken to be $(r, \theta) = (0.2\kpc, 0\degr) \rightarrow (800\kpc, 180\degr)$.
    A high-resolution uniform radial grid patch of $256$ cells covers the injection region ($r = 0.2 \rightarrow 2\kpc$, a resolution of $7\,\textrm{pc}$ per cell), while the remainder of the domain is uniformly covered by $1650$ cells.
    Azimuthally the jet injection region is uniformly resolved for each side by $300$ cells over the first $5\degr$, and $250$ cells over the next $12\degr$.
    A lower resolution uniform grid patch of $400$ cells is placed from $\theta = 17\degr \rightarrow 163\degr$, as for the adopted jet half-opening angle of $15\degr$ this region will mainly contain sideways expansion of the jet lobe and cocoon.
    At a distance of $200\kpc$ this provides a resolution of $58\,\textrm{pc}$ across the jet head.
    The lower radial boundary has an inflow condition within the jet nozzle; reflective boundary conditions were applied outside the jet nozzle and at the upper radial boundary.
    Axisymmetric boundary conditions were applied to both the lower and upper azimuthal boundaries.

    The jet injection region is defined in three dimensions as a sphere centred at the origin with radius $r_0 = 1\kpc$ (\autoref{fig:injection_zone}).
    Within this sphere the jet is injected as momentum and energy fluxes by overwriting the density, pressure, and velocity of these cells with the corresponding injection zone values, $(\rho, P, \vec{v}) = (\rho_\textrm{i}(r), P_\textrm{i}(r), \vec{v}_\textrm{i}(r))$.
    The jet density and pressure are defined in the injection sphere as 
    \begin{align}
      \rho_\textrm{i}(r) &= 2 \rho_\textrm{j} (1 + (r / r_0)^2)^{-1}\\
      P_\textrm{i}(r) &= 2^\Gamma P_\textrm{j} \left(\frac{\rho(r)}{2\rho(r_0)}\right)^\Gamma
      \,.
    \end{align}
    $P_\textrm{j}$ is tuned to keep perturbations from the injection zone small.
    The velocity in the injection region is defined along the jet propagation axis as $v_r = v_\textrm{j}$ if $\theta \le \theta_\textrm{j}$, and $0$ otherwise; here, $\theta_\textrm{j}$ is the half-opening angle of the jet.
    A passive tracer fluid is used to trace jet material; it is initialized to $0.0$ throughout the environment while given a value of $1.0$ inside the injection region.
    This fluid is then advected along with the jet flow and is used to quantify mixing between the jet and ambient material.
    At a radius of $r_0$, the edge of the jet injection zone is resolved by $6$ cells.
    In all simulations, the jet expands before collimation, so the number of cells across the collimated jet is larger than that of the injection region, providing sufficient resolution across the jet beam to capture recollimation dynamics.

    For each one of the two relativistic jets of a bipolar radio source, the total power is given \citep{Mukherjee2020} by
    \begin{equation}
      Q = \left[ \gamma (\gamma - 1) c^2 \rho_\textrm{j} + \gamma^2 \frac{\Gamma}{\Gamma - 1} P_\textrm{j} \right] v_\textrm{j} A_\textrm{j}\,,
      \label{eqn:rel_jet_power}
    \end{equation} where $\gamma = 1 / \sqrt{1 - v_\textrm{j}^2/c^2}$ is the bulk Lorentz factor of the flow, $c$ is the speed of light in a vacuum, $\Gamma$ is the adiabatic index, and $A_\textrm{j}$ is the cross-sectional area of the jet inlet.
    The jet density $\rho_\textrm{j}$ is calculated using Eq.~\ref{eqn:rel_jet_power} for a given combination of $Q,v_\textrm{j},P_\textrm{j},A_\textrm{j}$ as
    \begin{equation}
      \rho_\textrm{j} = \left( \frac{Q}{v_\textrm{j} A_\textrm{j}} - \gamma^2 \frac{\Gamma}{\Gamma - 1} P_\textrm{j} \right) \frac{1}{\gamma (\gamma - 1) c^2}\,.
      \label{eqn:rel_jet_density}
    \end{equation}
    When discussing relativistic fluid flows, it is useful to introduce a temperature parameter, $\Theta = P / (\rho c^2)$ \citep{Mignone2007a}, and in the case of relativistic jets specifically, $\chi$, the ratio between rest-mass energy and the thermodynamic part of the enthalpy \citep{Bicknell1994,Bicknell1995},
    \begin{equation}
      \chi = \frac{\Gamma - 1}{\Gamma} \frac{\rho_\textrm{j} c^2}{P_\textrm{j}}\,.
    \end{equation}
    Cold jets with $\Theta \ll 1$ ($\chi \gg1$) are dominated by kinetic energy, while hot jets with $\Theta \gg 1$ ($\chi \ll 1$) are dominated by thermal energy.
    When calculating the injection parameters of cold jets ($\chi \gg 1$), as is the case for the jets presented in this work, the ideal equation of state can be used \citep{Mignone2007a}.
    However, initially cold jet material will not remain kinetically dominated throughout the jet evolution due to shock-heating along the jet and at the hotspots.
    To account for this we use the Taub-Mathews (TM) \citep{Taub1948,Mathews1971,Mignone2007a} equation of state, which is built into \textsc{pluto} and handles both non-relativistic and relativistic gases.
    In the TM equation of state, the relativistic specific enthalpy is given as
    \begin{equation}
      h = \frac{5}{2} \Theta + \sqrt{\frac{9}{4} \Theta^2 + 1}\,,
    \end{equation}
    with an equivalent adiabatic index
    \begin{equation}
      \Gamma_\textrm{eq} = \frac{h - 1}{h - 1 - \Theta}\,.
    \end{equation}

    The relativistic generalisation of density contrast \citep{Marti1997,Krause2005,Bromberg2011} is defined as
    \begin{equation}
      \eta_\textrm{r} = \frac{\rho_\textrm{j} h_\textrm{j} \gamma^2}{\rho_\textrm{a} h_\textrm{a}}\,,
    \end{equation}
    with relativistic specific enthalpy $h = 1 + \Gamma e / \rho c^2$ for a cold jet given the equation of state for an ideal gas, $p = (\Gamma - 1)e$.
    $\eta_\textrm{r}$ can be compared with the non-relativistic jet density contrast $\eta$ such that for a light one-dimensional jet at velocity $v_\textrm{j}$, the hotspot would propagate at a velocity of $v_\textrm{HS} = \sqrt{\eta_\textrm{r}} v_\textrm{j}$.

  \subsection{Parameter study}\label{sec:parameter_study}

    We simulate a radio jet pair, each with a power of $Q=3 \times 10^{38}\,\textrm{W}$, typical of moderate-power FR IIs \citep{Turner2018,Shabala2020}.
    The main science simulations consist of a pair of relativistic jets on a three-dimensional grid.
    We vary the offset from the cluster centre ($0$ or $1$ core radii, corresponding to $0$ or $144\kpc$), as well as the inclination angle with respect to the centre ($0\degr$, $15\degr$, or $45\degr$).
    The jets are relativistic, $\gamma = 5$ ($v_j = 0.98 c$), have a half-opening angle $\theta_\textrm{j}=10\degr$, and are initially overpressured and underdense compared to the ambient medium.
    Finally, we conduct a resolution study with high-resolution two-dimensional relativistic simulations to verify that we are capturing the jet recollimation dynamics correctly in the three-dimensional simulations.

    The simulations and their parameters are listed in \autoref{tbl:simulations}.
    To aid with the discussion, in all offset cases we define the jet directed towards the cluster centre as the primary jet, while the jet directed away from the cluster centre as the secondary jet.

    The simulations presented here were carried out on the \textit{kunanyi} (Tasmanian Partnership of Advanced Computing) and \textit{Raijin} (National Computational Infrastructure) high-performance computing facilities.
    Between $1400$ and $4800$ cores were used, and the three-dimensional relativistic simulations each took approximately $300,000$ CPU hours.

    \begin{table*}
      \centering
      \caption{
        A list of parameters for the simulations presented in this paper.
        $Q$ is the total one-sided power of the radio source, which is the same for all runs.
        $v_\textrm{j}$ is the jet velocity,
        $r_\textrm{offset}$ is the radial offset from the environment centre,
        $\theta_\textrm{i}$ is the inclination angle of the jet with respect to the environment centre,
        $\Theta_\textrm{j}$ is the temperature parameter of the jet at injection,
        and $\eta_\textrm{r}$ is the relativistic generalised density contrast at injection.
      }
      \label{tbl:simulations}
      \begin{tabular}{ l l r r r r r r}
        \hline
                            & Code         & $Q$ (W)           & $v_\textrm{j}$ ($c$) & $r_\textrm{offset}$ (kpc) & $\theta_\textrm{i}$ (\degr) & $\Theta_\textrm{j}$ & $\eta_\textrm{r}$  \\
        \hline
        Main simulations    & off0r        & $3\times 10^{38}$ & $0.98$               & $0$                       & $0$                         & $3.2\times 10^{-3}$ & $5.9\times10^{-2}$ \\
                            & off1r        &                   &                      & $144$                     & $0$                         &       & $8.8\times10^{-2}$ \\
                            & off1r-15deg  &                   &                      &                           & $15$                        &        &                    \\
                            & off1r-45deg  &                   &                      &                           & $45$                        &        &                    \\
        2D resolution study & 2doff0r      &                   & $0.90$               & $0$                       & $0$                         & $7.5\times 10^{-4}$  & $3.7\times10^{-1}$ \\
                            & 2doff1r      &                   &                      & $144$                     & $0$                         & $5.1\times 10^{-4}$  & $5.5\times10^{-1}$ \\
        \hline
      \end{tabular}
    \end{table*}

  \subsection{Synthetic radio observables}\label{sec:synth_radio}

    Emission due to synchrotron radiation is the main method by which the large-scale structure of radio sources can be observed.
    Synchrotron radiating particles are likely accelerated both along the jet and at the jet hotspots, continuing to radiate as they flow back into the radio lobes \citep{Hardcastle2016,Matthews2020,Rieger2021}. 
    Thus to compare observed radio source structure with simulations, the synchrotron emissivity of a simulated radio source needs to be calculated.
    While the simulations presented here do not contain the magnetic fields necessary for a complete treatment of synchrotron radiation, the emissivity can be calculated by assuming a constant departure from equipartition for the magnetic field and electron energy densities \citep{Kaiser1997a,Turner2015}.
    Magnetohydrodynamic simulations show that the ratio of particle to magnetic pressure (plasma $\beta$) remains on average constant over time \citep{Hardcastle2014}, but has large spread spatially due to lobe turbulence \citep{Gaibler2009}.
    Thus our assumption has the effect of smoothing out any local magnetic field variations across the lobes.

    The synchrotron emissivity of a packet of electrons with a power-law distribution of electron energies $N(E) = \kappa E^{-q}$ emitting at an angular frequency $\omega$ is given by \citet{Longair2011} as
    \begin{equation}
      j(\omega) = A \frac{\sqrt{3 \pi} e^3 B}{16 \pi^2 \epsilon_0 c m_e (q+1)} \kappa \left( \frac{\omega m_e^3 c^4}{3 e B} \right)^{- \frac{q-1}{2}}\,.
      \label{eqn:volume_emissivity_longair}
    \end{equation}
    The local magnetic field strength $B$ is related to the local electron density $u_\textrm{e}$ through a constant departure from equipartition, $f_\textrm{B} = u_\textrm{B} / u_\textrm{e}$.
    \citet{Kaiser1997a} present a relationship between cocoon pressure and energy densities, $P=(\Gamma_c-1)(u_\textrm{e}+u_\textrm{B}+u_\textrm{T})$, which is used to relate the local pressure to the local electron density, assuming a negligible thermal energy component.
    The electron distribution normalisation $\kappa$ depends on observationally-informed distribution properties (electron Lorentz factor limits $\gamma_\textrm{min}$ and $\gamma_\textrm{max}$, energy power-law slope $q$), and local pressure $P$.
    The synchrotron emissivity for a given frequency $\nu$ and local pressure $P$ can then be written as
    \begin{equation}
      j_\nu = j_0 P^\frac{q + 5}{4}\,,
      \label{eqn:volume_emissivity_pressure}
    \end{equation}
    where the full derivation for $j_0$ in units of W Hz$^{-1}$, the volume emissivity coefficient, is given in \citet{Yates2018}.

    Spectral ageing is not included in this work, so the synthetic radio spectrum has an identical shape for all frequencies; frequency serves only to change the normalisation $j_0$.
    The emissivity is weighted by the jet tracer fluid, to account for mixing of jet and ambient material.
    We note that this method for calculating the emissivity does not include any losses, which alter the observed morphology of the radio source \citep{Turner2018a} and the evolution of luminosity with time \citep{Kaiser1997a,Shabala2013a,Turner2015,Hardcastle2018}.
    In this work, we focus on the dynamics of the radio source, and so neglect both radiative and adiabatic losses, in addition to Doppler boosting due to relativistic fluid motions.
    \citet{English2016} showed that the effect of Doppler boosting on total luminosity is small due to both the small amount of emission associated with the jets and the relatively low bulk Lorentz factors found on the grid, consistent with our simulations.

    Using Eq.~\ref{eqn:volume_emissivity_pressure}, the emissivity per unit volume for each simulation grid cell can be calculated.
    We take the electron energy power-law slope to be $q = 2.2$ as in \citep{Hardcastle2013}, corresponding to a spectral index $\alpha=-0.6$ typical of radio lobes; the ratio of magnetic to particle energy densities as $f_\textrm{B} = 0.1$; and the minimum and maximum electron Lorentz factors as $\gamma_\textrm{min}=10$ and $\gamma_\textrm{max}=10^{5}$.
    The total lobe luminosity $L_\nu$ is an integral of emissivity over the lobe volume,
    \begin{equation}
      L_\nu = \int j_\nu\,\mathrm{d}V\,.
    \end{equation}
    The two-dimensional surface brightness can be obtained by integrating along a chosen line of sight $r$,
    \begin{equation}
      B_\nu = \frac{1}{4 \pi} \int j_\nu (r)\,\mathrm{d}r\,.
    \end{equation}
    Following \citet{Yates2018}, we place our simulated radio sources at $z=0.05$, interpolate the surface brightness maps onto a pixel size of $1.8\,\textrm{arcsec}^2$, and approximate the effects of an observing beam by convolving the surface brightness maps with a two-dimensional circular Gaussian beam with a full width at half maximum of $5\,\textrm{arcsec}$.
    The observing frequency is chosen to be $\nu = 1.4\,\textrm{GHz}$.
    These parameters are characteristic of the VLA FIRST \citep{Becker1995} survey, and similar to ongoing and future mid-frequency surveys by SKA pathfinders.
    The radio sources are placed in the plane of the sky so that the line-of-sight integral is performed along the x-axis.
    Our synthetic emissivity model captures the hotspot and recently shocked lobe emission with reasonable accuracy, but also produces equatorial emission which is not observed in real radio sources due to electron aging; a detailed discussion of this point is deferred to Yates-Jones et al. (in preparation).

\section{Results}\label{sec:results}

  \subsection{Dynamics and morphology}
  \label{sec:dynamics_and_morphology}

    \begin{figure*}
      \includegraphics{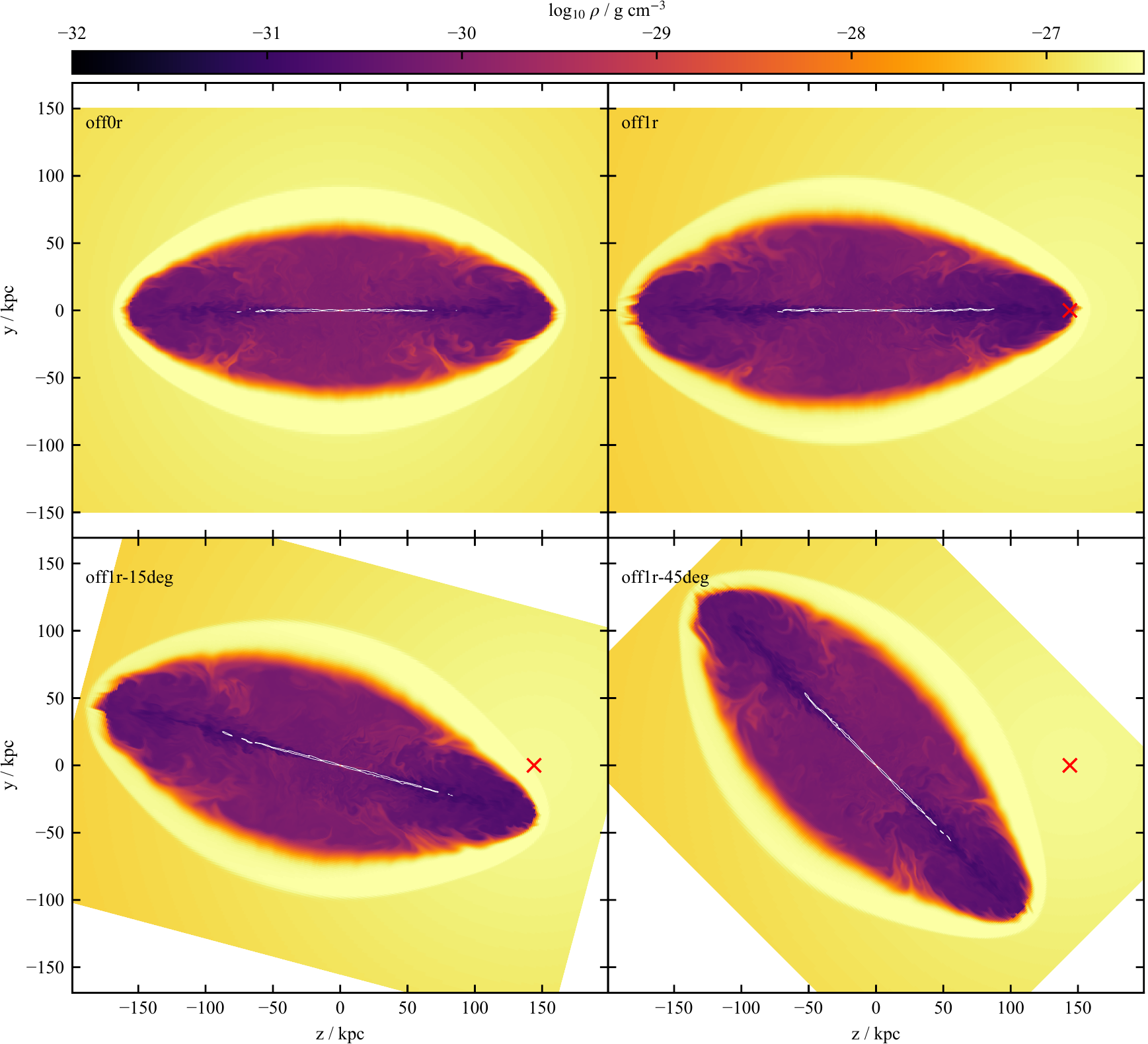}
      \caption{
        Maps of midplane density for the suite of relativistic three-dimensional simulations at $t=32\myr$.
        The slices are taken in the Y-Z plane.
        The upper left panel shows the jets in a non-offset environment, while the remaining three panels show jets in an environment offset by $1$ core radius ($144\kpc$).
        The jets in the two lower panels are inclined $15\degr$ and $45\degr$ away from the environment centre respectively; the image panels are rotated by the corresponding amount such that the environment centre is aligned with the horizontal axis.
        The location of the environment centre is marked in the three offset simulations with a red cross.
        White contours on each plot denote an internal Mach number of $5$.
        The primary jet is propagating towards the environment centre, in the $+$Z direction, while the secondary jet is propagating in the $-$Z direction.
        In the offset environments, the primary jet is propagating into a rising density and pressure profile, and produces a shorter, narrower cocoon, compared to the secondary jets.
        Lateral asymmetry is observed across the primary lobe when the jet is inclined with respect to the environment centre.
        In this case the jet head preferentially expands away from the dense centre of the environment creating a lopsided structure.
      }
      \label{fig:density_map_relativistic}
    \end{figure*}

    \begin{figure*}
      \includegraphics{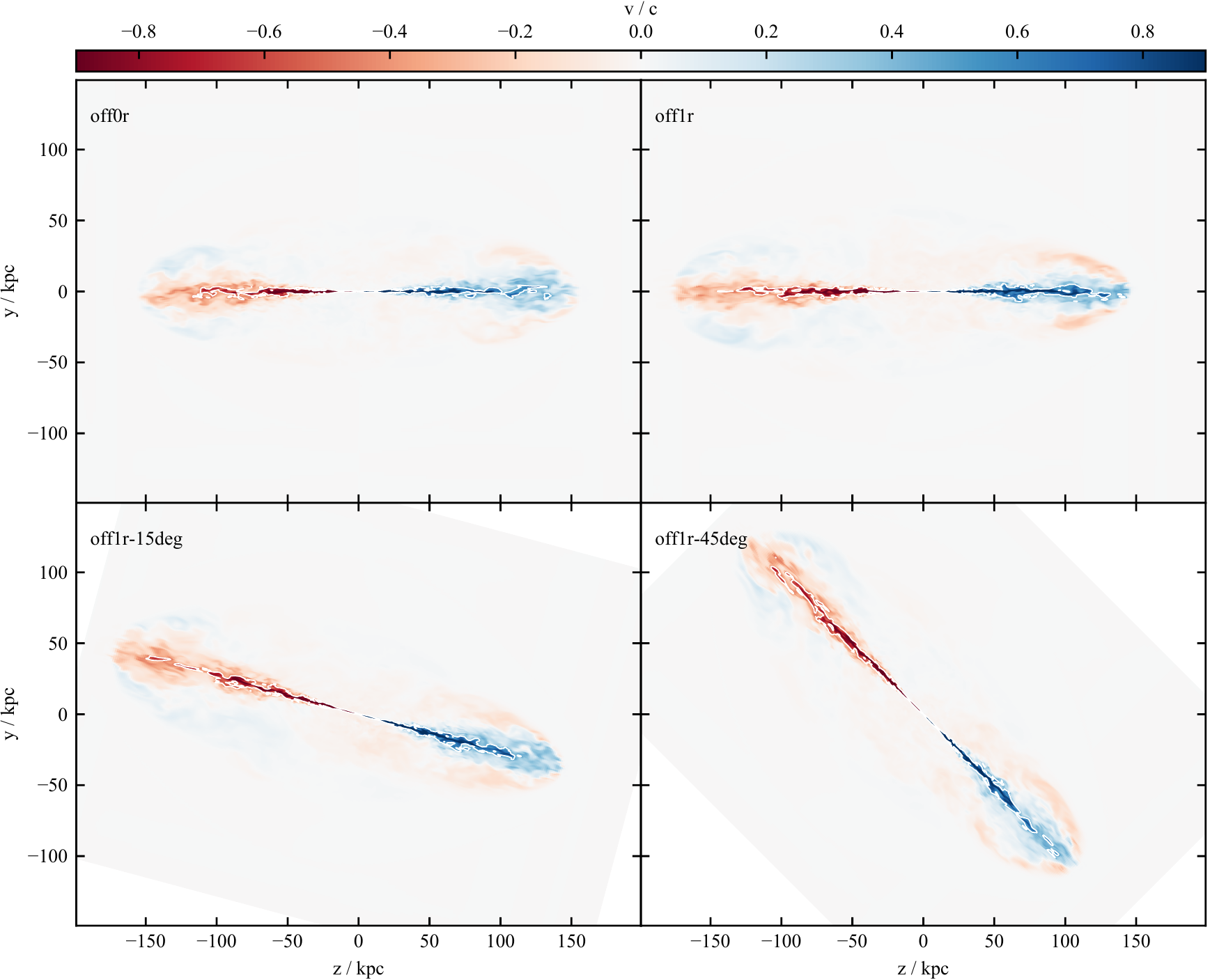}
      \caption{
        Maps of midplane velocity along the jet axis for the suite of relativistic three-dimensional simulations at $t=32\myr$.
        The slices are taken in the Y-Z plane, oriented as in \autoref{fig:density_map_relativistic}.
        White contours on each plot denote a bulk velocity magnitude of $0.5c$.
      }
      \label{fig:velocity_map_relativistic}
    \end{figure*}

    \begin{figure*}
      \includegraphics{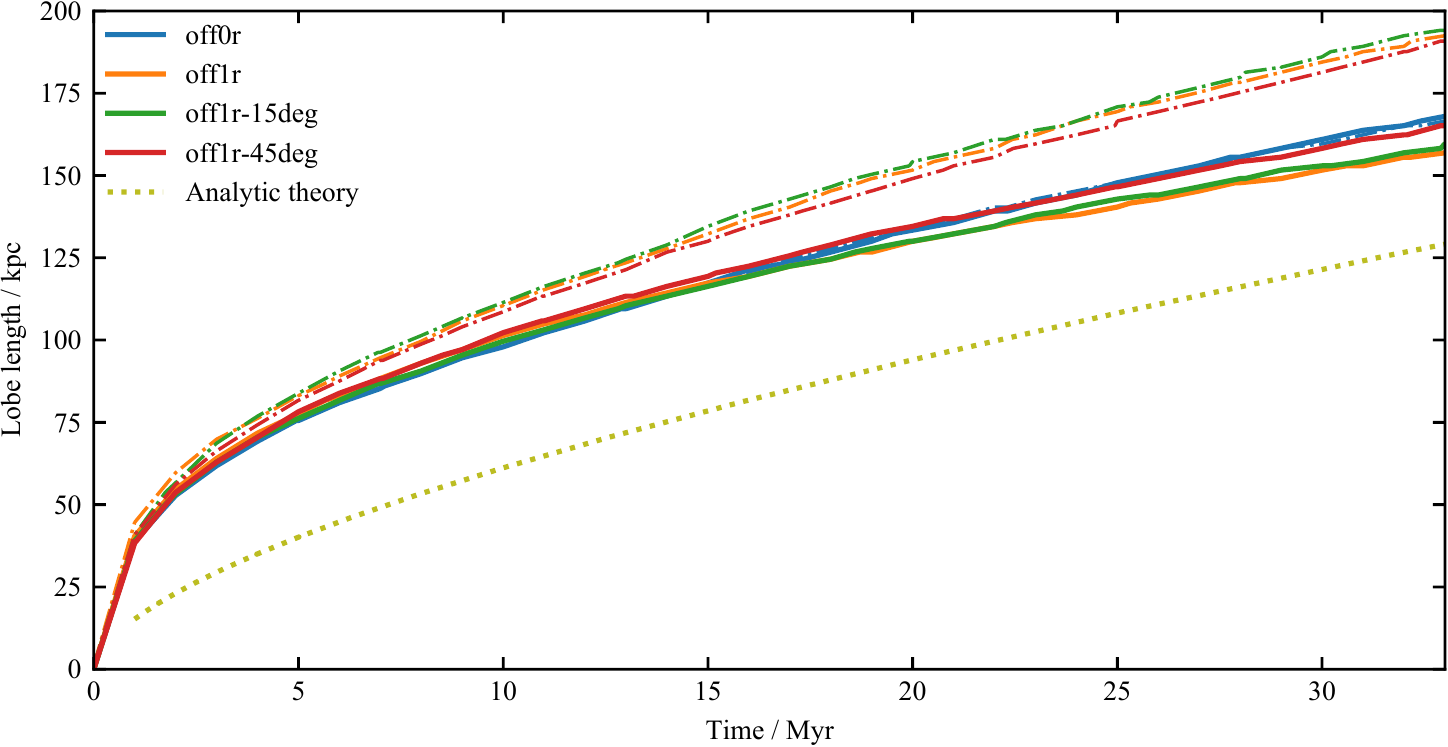}
      \caption{
        Lobe length as a function of time for all three-dimensional simulations, calculated using a jet tracer threshold.
        The primary lobe is shown as the solid line, while the secondary lobe is shown as the dot-dashed line.
        The theoretical evolution of a non-offset lobe modelled using the RAiSE analytic model \citep{Turner2015} with the same jet parameters and environment is shown as the dashed line.
      }
      \label{fig:lobe_length_evolution}
    \end{figure*}

    \begin{figure*}
      \includegraphics{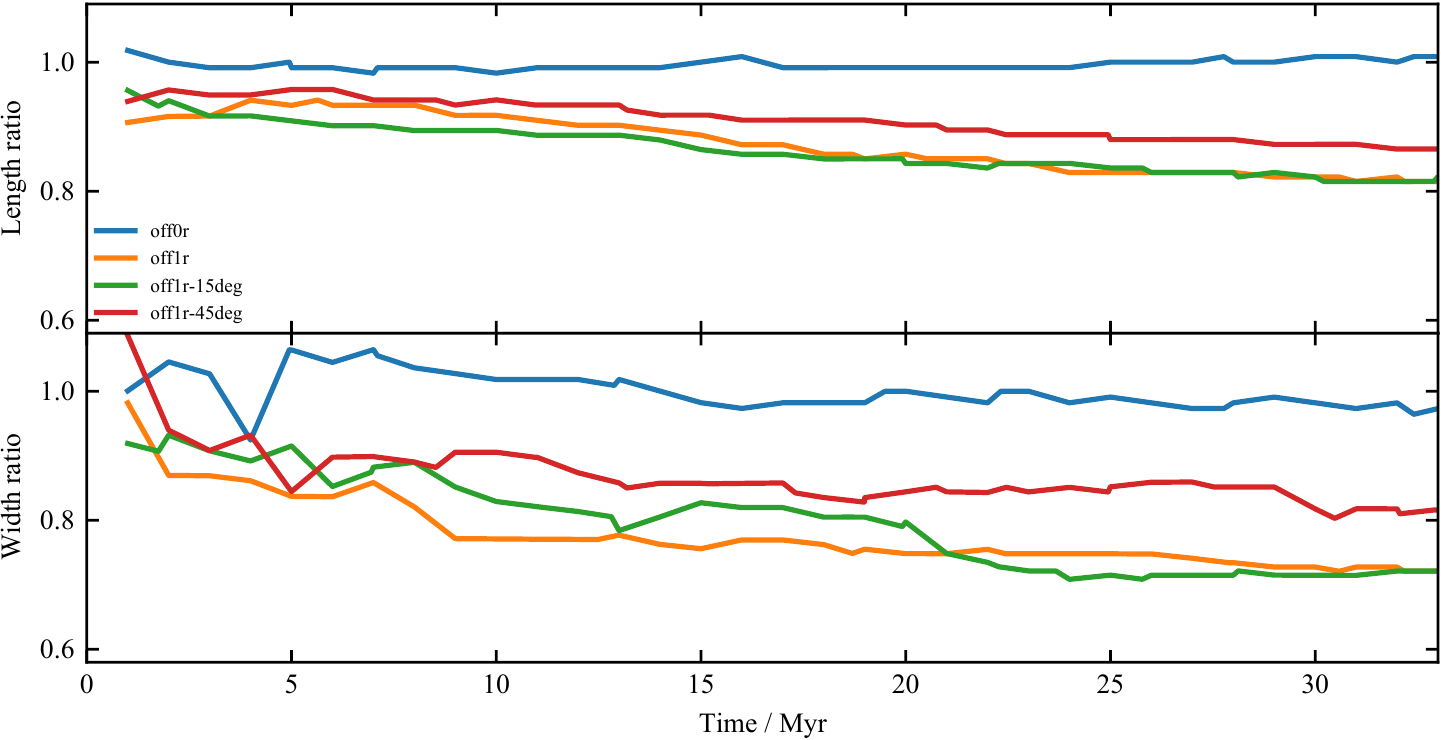}
      \caption{
        Lobe length (upper) and width (lower) ratios (primary / secondary) as a function of time for all three-dimensional simulations.
      }
      \label{fig:lobe_length_width_ratio_evolution}
    \end{figure*}

    \begin{figure*}
      \includegraphics{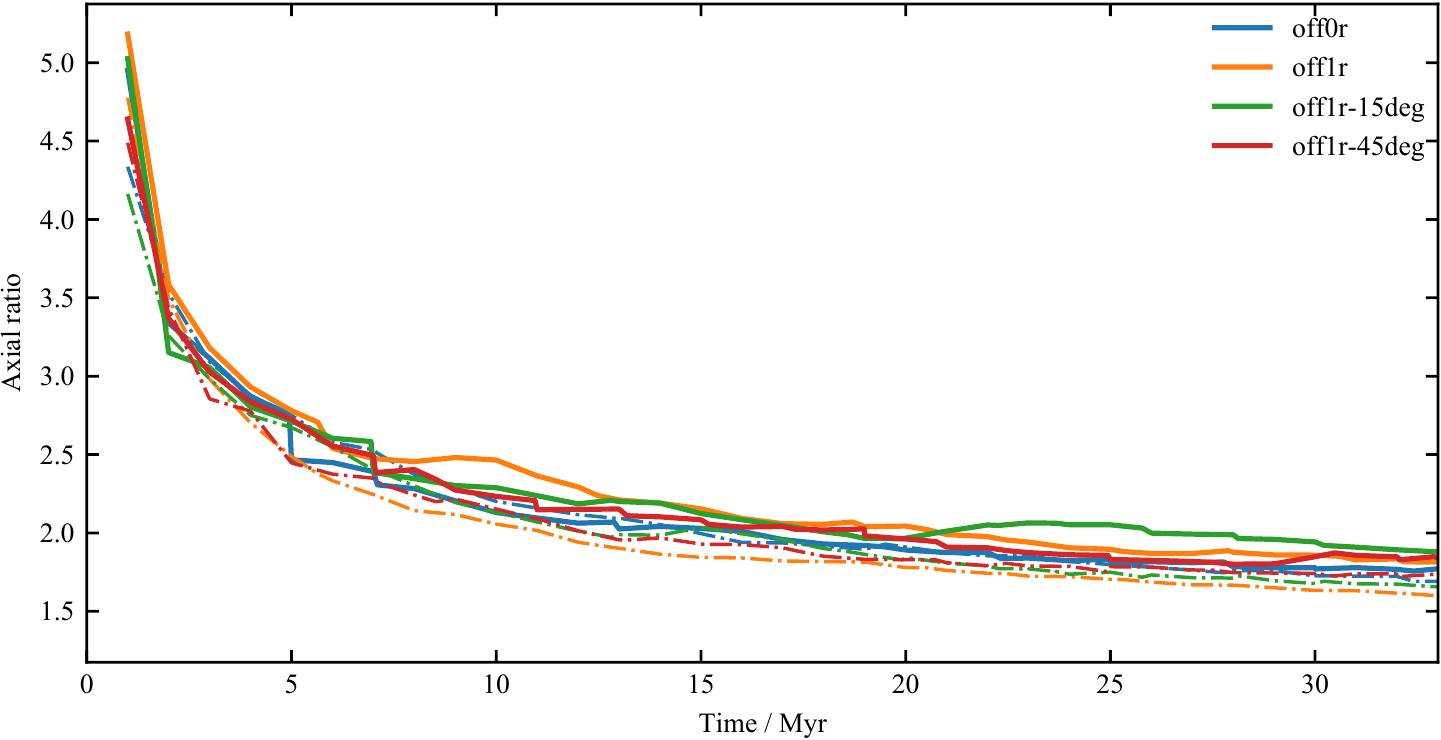}
      \caption{
        Lobe axial ratio as a function of time for all three-dimensional simulations.
        Line styles are as in \autoref{fig:lobe_length_evolution}.
      }
      \label{fig:lobe_axial_ratio_evolution}
    \end{figure*}

    We show logarithmic density maps of our main simulations at the end of the simulation time in \autoref{fig:density_map_relativistic}.
    The jets re-collimate successfully before entraining gas later via three-dimensional instabilities and transitioning to turbulence.
    The jet disruption is very similar to the mechanism described in \citet{Massaglia2016}, and is essentially due to the low Mach number in the collimated jet.
    When the jet first collimates, the internal relativistic Mach number is above $10$.
    Along the jet, it then oscillates between approximately $6$ and $10$, with a declining trend of the minima.
    Where it drops below $5$ (shown as the white contour in \autoref{fig:density_map_relativistic}), the jet disrupts.
    We find that the jet disruption radius is roughly equal for a primary/secondary jet pair in a given environment over the simulated time.
    Additionally, no significant differences are present between environments.

    In \autoref{fig:velocity_map_relativistic} we show midplane slices of velocity along the jet axis.
    The white contours (which denote a bulk velocity of $0.5c$) highlight the deceleration of jet material after the disruption point.
    Clear backflow is present in all simulations, and slight asymmetries are introduced in the inclined jets due to the environment (the lower two panels of \autoref{fig:velocity_map_relativistic}). 
    The overall lobe shape is pinched in the case of the primary jet (see, e.g., top-right panel of the same figure).

    We measure the lobe morphology using a tracer threshold of $10^{-6}$.
    \citet{Gaibler2009} have shown that lobe morphologies do not strongly depend on the exact value of this parameter.
    \autoref{fig:lobe_length_evolution} shows the lobe length evolution as a function of time.
    We also plot the lobe length evolution as modelled by RAiSE \citep{Turner2015}, using the same jet parameters and environment.
    The jets undergo a fast expansion phase for the first few Myrs, before beginning to slow; this initial `breakout' phase is not captured by the RAiSE model.
    However, at later times, the rate of jet length evolution is consistent with the analytical model.

    In all offset cases, the primary jet is shorter---as predicted by analytic theory \citep[e.g.][]{Kaiser1997} for a rising density profile---and the lobe is pinched compared to the secondary jet.
    This causes the observed lobe length asymmetry; initially, the primary and secondary lobes are similar, but they diverge as the source ages.
    Inclining the jet with respect to the cluster centre has no significant effect on lobe length.

    In \autoref{fig:lobe_length_width_ratio_evolution} we plot the evolution of the length and width ratios for all three-dimensional simulations.
    All offset simulations show significant departures from the baseline, demonstrating that the secondary lobe is becoming both longer and wider than the primary with time.

    Following \citet{Hardcastle2013} we define the lobe axial ratio as the ratio between the lobe length and the lobe width, as measured at the midpoint of the lobe; the axial ratio for all simulations is plotted in \autoref{fig:lobe_axial_ratio_evolution}.
    The lobes produced by relativistic jets are initially elongated with high axial ratios, during the breakout phase of the jet.
    As the jet head propagation slows, the lobes begin to inflate because they are overpressured compared to the environment, causing the axial ratio to approach the range $[1.5, 2]$.
    This is significantly lower than in the non-relativistic simulations of \citet{Hardcastle2013}.
    Since a large fraction of observed axial ratios is between $1.5$ and 2 \citep[Figs 6-9]{Mullin2008}, our relativistic jets with self-consistent hydrodynamic collimation may explain many radio sources better than non-relativistic ones, although projection effects non-trivially affect this.

    In contrast to the length and width ratio evolution, the difference in axial ratio between the primary and secondary lobes for the $1\rcore$-offset relativistic simulations is small.
    At later times this difference becomes more pronounced, with the primary lobes having a consistently higher axial ratio compared to the secondary lobes.
    The width ratio evolution is a function of both the length ratio (as the lobe width is measured at the midpoint between the core and the hotspot), as well as the pinching occurring in the primary lobes.

    Relativistic jets produce initially narrow lobes.
    The length evolution of the lobe is driven by balancing both the jet head pressure and momentum flux against the ambient density and pressure, while the transverse expansion is determined solely by the lobe pressure \citep{Hardcastle2013}.
    All simulations have axial ratios that evolve with time, indicating that they are not expanding self-similarly, consistent with analytical models of \citet{Turner2015,Hardcastle2018}.

  \subsection{Simulated radio emission}
  \label{sec:observable_properties}

    \begin{figure*}
      \includegraphics{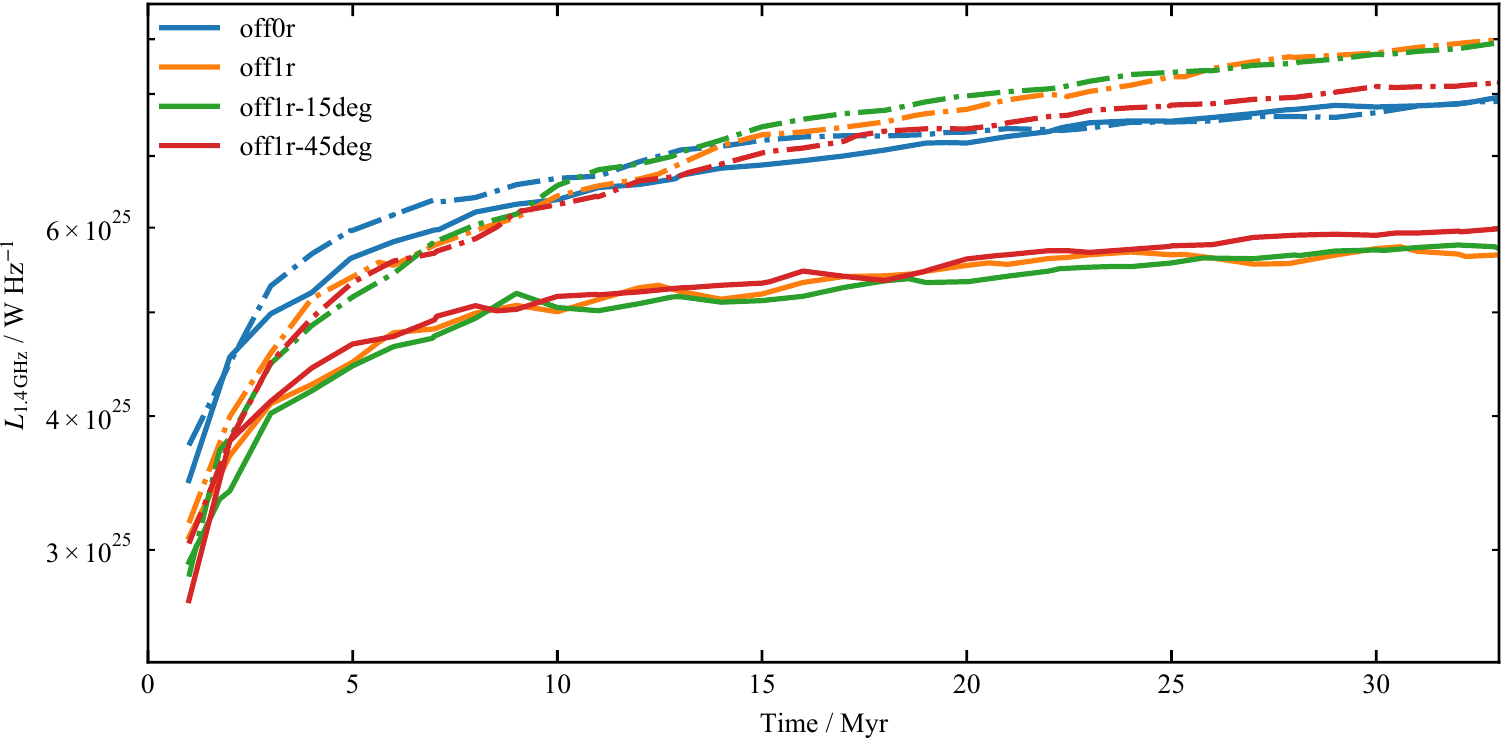}
      \caption{
        Luminosity evolution as a function of time for all three-dimensional simulations.
        The luminosity is calculated for an observing frequency of $1.4\,\textrm{GHz}$ by integrating the tracer-weighted emissivity over the simulation domain.
        No losses are included in the emissivity calculation, so these are the upper limits of the luminosity.
        Line styles are as in \autoref{fig:lobe_length_evolution}.
      }
      \label{fig:lobe_luminosity_evolution}
    \end{figure*}

    \begin{figure*}
      \includegraphics{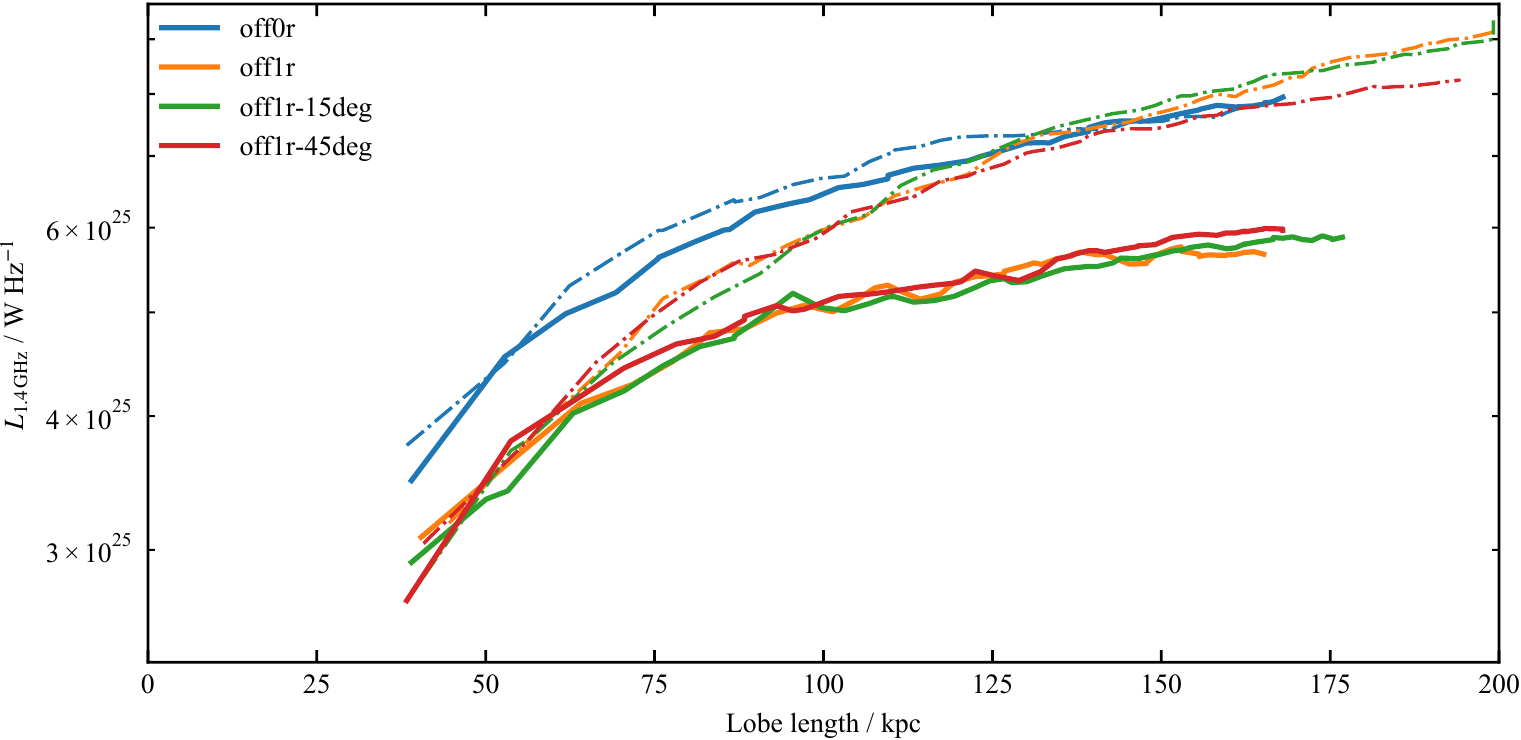}
      \caption{
        Evolutionary tracks through the size-luminosity diagram for all three-dimensional simulations.
        Luminosity is calculated as for \autoref{fig:lobe_luminosity_evolution}.
        Line styles are as in \autoref{fig:lobe_length_evolution}.
      }
      \label{fig:pd_tracks_3D}
    \end{figure*}

    \begin{figure*}
      \includegraphics{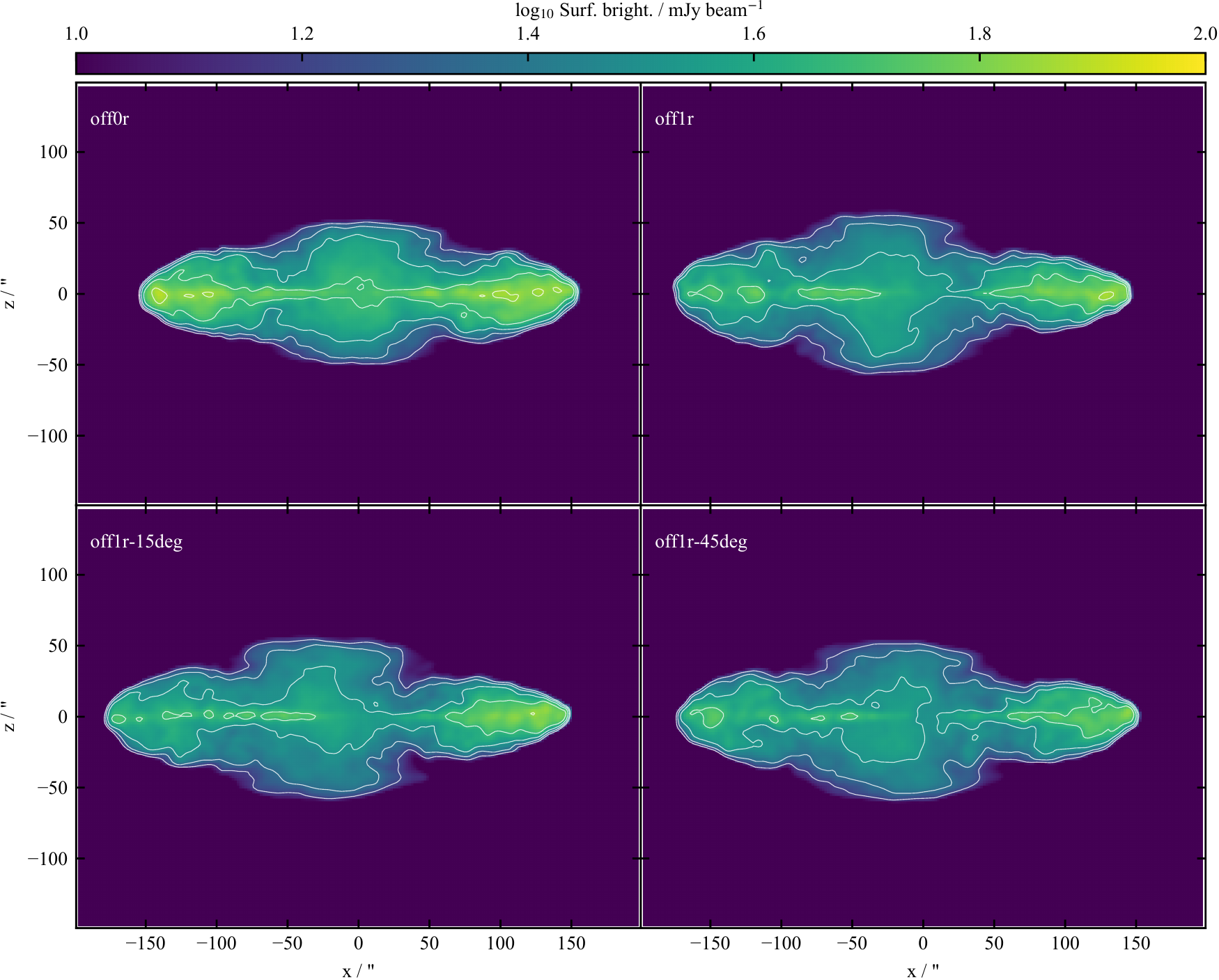}
      \caption{
        Synthetic surface brightness maps for the relativistic three-dimensional simulations at $32\myr$.
        The surface brightness is calculated for an observing frequency of $1.4\,\textrm{GHz}$, at a redshift of $z=0.05$, observed with a 2D Gaussian beam with $\theta_\textrm{FWHM} = 5\,\textrm{arcsec}$ and a pixel size of $1.8\,\textrm{arcsec}^2$.
        The surface brightness is shown in units of mJy beam$^{-1}$, and there are $6$ evenly spaced contours in log space from $\log_{10} \textrm{Surf. bright.} = 1.5\,\textrm{to}\,2.5\,\textrm{mJy beam}^{-1}$ 
        The primary lobe exhibits enhanced surface brightness at the hotspot, and reflects the narrower underlying cocoon.
      }
      \label{fig:surface_brightness_map_relativistic}
    \end{figure*}

    In addition to the purely morphological differences, the primary and secondary jets have different lossless lobe luminosities and size-luminosity tracks, as shown in \autoref{fig:lobe_luminosity_evolution} and \autoref{fig:pd_tracks_3D}.
    The secondary jet has consistently higher lobe luminosities for a given time, due to the increased volume.
    Lobe pressure can be considered homogeneous across the entire radio source due to the high sound speed, except for the overpressured hotspots.
    This results in a strong correlation between luminosity and lobe volume for the lossless calculations used in this work, and so the secondary jets produce lobes with higher luminosities due to their larger volume.
    The luminosity evolution of all our primary lobes is very similar; the same is true for the luminosity evolution of all our secondaries.
    Hence, we conclude that our results are robust for variations of the angle the jet makes with the environment symmetry axis.

    \autoref{fig:pd_tracks_3D} shows the impact of environment profiles on size-luminosity tracks.
    The primary PD tracks of the offset simulations begin to flatten out after the initial $\sim 80 \kpc$, as the environment itself starts to flatten out.
    The secondary PD tracks on the other hand continue rising for the $1\rcore$-offset simulations inclined at $0$ and $15\degr$.

    We show synthetic radio images in \autoref{fig:surface_brightness_map_relativistic}.
    We note once more that radiative losses would make the central parts of the lobes less prominent and increase the surface brightness contrast between the hotspot and lobe.
    Despite the jets becoming unstable and undergoing entrainment, the general source structure is FR II-like and clear hotspots are produced.
    Subtle differences are consistently visible between the primary and secondary lobes.
    The primary jets exhibit enhanced surface brightness at the hotspots as expected from theory \citep{Kaiser1997a}.

\section{Comparison to observations}\label{sec:obs}

  \begin{figure*}
    \includegraphics{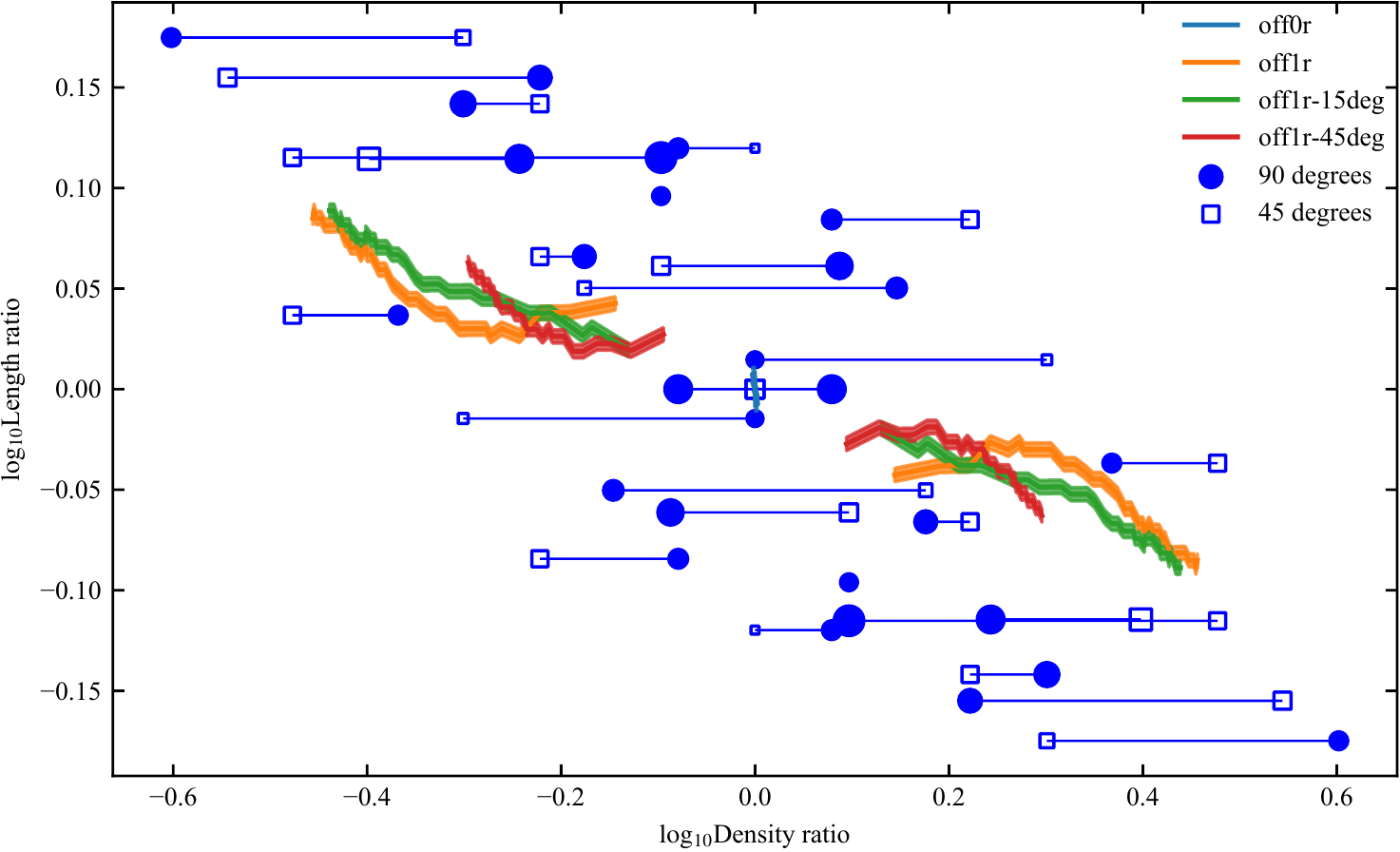}
    \caption{
      Length asymmetry ratio as a function of density asymmetry ratio using data courtesy of \citet[Figure 9]{Rodman2019}.
      The numerical simulation data for the three-dimensional simulations is shown in the same colours as \autoref{fig:lobe_length_evolution}, from $t=0$ to $t=32\myr$.
      Error bands are shown for the simulation tracks, taken to be the $\pm1\sigma$ bounds of the $0\rcore$-offset simulation.
      Observed data is represented as the closed circles and open squares, where the symbol size is proportional to the galaxy count.
      The observed data has been mirrored along the line $y=x$, to reflect the freedom in choosing the primary radio lobe.
      Optical asymmetry ratio calculated using a cone of $90\degr$ is shown as the closed circles, while a cone of $45\degr$ is shown as the open squares---the observed data for each radio galaxy is connected by a horizontal line.
    }
    \label{fig:asymmetry_ratio}
  \end{figure*}

  A large number of extended radio sources show some degree of length asymmetry in their radio lobes.
  A large analysis of these asymmetric objects was presented by \citet{Rodman2019}.
  The authors used a sample of mainly double-lobed radio sources identified as having asymmetric, straight radio lobes through the Radio Galaxy Zoo citizen science project \citep{Banfield2015}, using data from the Faint Images of the Radio Sky at Twenty Centimeters \citep[FIRST;][]{Becker1995} and the Australian Telescope large Area Survey \citep[ATLAS;][]{Norris2006,Middelberg2008}.
  Their final sample contained radio sources with redshift $z < 0.3$, selected to have approximately straight radio lobes greater than $100\kpc$ in size, with at least 20 neighbour galaxies associated through photometric redshifts to the host galaxy using the Sloan Digital Sky Survey (SDSS) DR10 \citep{Ahn2014}.
  Galaxies were associated with one of the radio lobes based on whether they lay within either a $45$ or $90\degr$ cone aligned with the lobe major axis.

  The key finding of \citet{Rodman2019} is a strong anti-correlation between the radio lobe length and number density of galaxies associated with that lobe.
  Use of optical galaxy clustering as a good proxy for the underlying environment is supported by \citet{Schindler1999}, who showed that the galaxy and intra-cluster gas distributions in the Virgo cluster are similar.
  A length-environment density anti-correlation is predicted by \citep{Kaiser1997,Turner2015}, and \citet{Rodman2019} concluded that large-scale environment asymmetry is driving the radio lobe asymmetry.

  In \autoref{fig:asymmetry_ratio} we compare our simulations with data from Fig. 9 from \citet{Rodman2019}.
  Those authors plot the lobe length asymmetry of observed radio sources against the environment asymmetry as derived from optical galaxy clustering.
  To reproduce their plot, the lobe length asymmetry ratio is calculated at each timestep as $l_\textrm{ratio} = l_\textrm{p} / l_\textrm{s}$ and plotted against the environment density ratio $\rho_\textrm{ratio} = \rho(l_\textrm{p}) / \rho(l_\textrm{s})$ for each simulation.
  The initial environment density (given by Eq.~\ref{eqn:density-isothermal-beta}) is used, and $l_\textrm{p}$ and $l_\textrm{s}$ refer to the primary and secondary lobe lengths respectively.
  We plot the environment density ratio on the same axis as the optical galaxy clustering asymmetry ratio.

  We find that the simulations reproduce the observed anti-correlation found by \citet{Rodman2019}, and fill a significant portion of observed parameter space.
  As the environment density asymmetry increases with source length (and hence age), these tracks show the temporal evolution of a radio source through this diagram.
  An estimate of the scatter in our method is given by the asymmetry tracks for the $0$-offset simulations, which is confined to the very centre of the parameter space.
  While our simulations do not completely explain the observed asymmetries, they do reproduce the overall trend.
  Based on this, our results imply that a significant portion of the observed length asymmetry in radio lobes can be traced back to large-scale environment asymmetries, in agreement with \citet{Rodman2019}.

\section{Discussion}
\label{sec:discussion}

  It is important to consider the assumptions present in our simulations before summarizing our results.
  These simulations do not include magnetic fields; this is a valid assumption given their relatively small influence on the large-scale lobe morphology.
  However, magnetic fields are likely to play a role in jet stability and collimation \citep{Matsumoto2021}, something which we aim to investigate in future work.
  The absence of magnetic fields also necessitates an approximate radio emissivity model.
  The main effect of losses is to dim radio lobes in the central parts, where electrons from the oldest part of the backflow reside.
  Since our main interest here is the lobe morphology towards the tips of the lobes, the basic lossless model used in this work is sufficient for discussing the observable large-scale morphological differences.

  Interestingly, we find that towards the end of the simulations, our jets are unstable on the 10s of kpc scale, despite having a Lorentz factor of 5.
  We have convinced ourselves in two-dimensional axisymmetric control runs that this is purely a three-dimensional (and therefore realistic) effect.
  This is reminiscent of the three-dimensional jet disruption for low Mach number jets demonstrated in much detail by \citet{Massaglia2016}.
  The interesting corollary here is that stable FR II jets with our chosen power (intermediate for FR IIs) require Lorentz factors higher than 5 to form a stable jet out to the 100 kpc scale.
  This finding is independent of the environments simulated, as jet stability is determined by the cocoon properties (density, pressure) which are similar across the different simulations.

  Nevertheless, our sources have clear FR II structure from our simulated radio maps and hence the comparison to the observations is justified.
  As in the observations, we find that environment asymmetry translates to lobe length asymmetry.
  Our simulations take full account of the hydrodynamic recollimation of the relativistic jet.
  Hence, the system is free to adjust jet recollimation, and hence jet width dynamically with the lobe pressure, and our jets stay collimated for long enough, for that process to be simulated faithfully.
  Our central lobe pressures are, however, uniform enough, such that the recollimation happens similarly for both jets of a radio source.
  Hence, the lobe asymmetry follows the naive expectation.
  Significant random scatter may be added to the lobe length asymmetry from the initial interactions of the jets with the dense, stochastically clumped, interstellar medium within the galaxy \citet{Gaibler2011a}.

\section{Summary and conclusions}\label{sec:conclusions}
    
  In this work, we have presented three-dimensional simulations of powerful relativistic radio jets in an isothermal beta environment, to study the effect of asymmetric environments on large-scale lobe morphology and radio observables.
  These jets have varying offsets and inclination angles with respect to cluster centre; all other jet parameters are held constant.
  We follow their evolution from conical injection and subsequent collimation to inflation of the large-scale radio lobes. 
  In summary, our work has shown the following.
  \begin{enumerate}
    \item Lobes propagating into denser environments are consistently shorter, exhibit a pinched morphology, and have enhanced surface brightness at the hotspot when compared with their counterparts.
    \item The inflated cocoon exhibits morphology deviations with inclination angle; backflow at the primary jet head preferentially expands away from the environment centre due to the increased density.
      The effect is however rather small, and overall deviations from axisymmetry appear to be dominated by random effects of three-dimensional instabilities.
      The backflow is consistently narrower in the denser environment.
    \item The jet collimation process is similar for both jets of a radio source regardless of offset and inclination angle.
      This is due to the homogeneous pressure distribution in the cocoon.
    \item Our simulations reproduce the observed link between radio source asymmetry and environment asymmetry \autoref{fig:asymmetry_ratio}.\\
      This supports the theory that jet-environment interaction plays a significant role in determining radio source morphology at large scales and opens the possibility of using radio sources as probes of the host environment in observations by next-generation survey instruments like SKA and it's pathfinders.
  \end{enumerate}

\section*{Acknowledgements}

  We thank Aikaterini Vandorou, Geoffrey Bicknell, Dipanjan Mukherjee, and Alex Wagner for useful discussions.
  We also thank an anonymous referee for their useful comments.
  PYJ thanks the University of Tasmanian for an Australian Postgraduate Award, the ARC Centre of Excellence for All Sky Astrophysics in 3 Dimensions for a stipend, and both the University of Tasmania and the Astronomical Society of Australia for their international travel support.
  SS thanks the Australian Government for an Endeavour Fellowship 6719\_2018.
  PYJ and SS thank the Centre for Astrophysics Research at the University of Hertfordshire for their hospitality\\
  \noindent This work was supported by resources awarded under Astronomy Australia Ltd’s ASTAC merit allocation scheme, with computational resources provided by the National Computational Infrastructure (NCI), which is supported by the Australian Government.
  We also gratefully thank the Tasmanian Partnership for Advanced Computing of the University of Tasmania for the computational resources provided.
  We acknowledge the work and support of the developers providing the following Python packages: Astropy \citep{AstropyCollaboration2018,AstropyCollaboration2013}, JupyterLab \citep{Jupyter}, Matplotlib \citep{Matplotlib}, NumPy \citep{NumPy}, and SciPy \citep{SciPy}.

\section*{Data Availability}

  The data underlying this article will be shared on reasonable request to the corresponding author.

  %%%%%%%%%%%%%%%%%%%%%%%%%%%%%%%%%%%%%%%%%%%%%%%%%%


  \bibliographystyle{mnras}
  \bibliography{asymmetry-local.bib}

\begin{thebibliography}{}
\makeatletter
\relax
\def\mn@urlcharsother{\let\do\@makeother \do\$\do\&\do\#\do\^\do\_\do\%\do\~}
\def\mn@doi{\begingroup\mn@urlcharsother \@ifnextchar [ {\mn@doi@}
  {\mn@doi@[]}}
\def\mn@doi@[#1]#2{\def\@tempa{#1}\ifx\@tempa\@empty \href
  {http://dx.doi.org/#2} {doi:#2}\else \href {http://dx.doi.org/#2} {#1}\fi
  \endgroup}
\def\mn@eprint#1#2{\mn@eprint@#1:#2::\@nil}
\def\mn@eprint@arXiv#1{\href {http://arxiv.org/abs/#1} {{\tt arXiv:#1}}}
\def\mn@eprint@dblp#1{\href {http://dblp.uni-trier.de/rec/bibtex/#1.xml}
  {dblp:#1}}
\def\mn@eprint@#1:#2:#3:#4\@nil{\def\@tempa {#1}\def\@tempb {#2}\def\@tempc
  {#3}\ifx \@tempc \@empty \let \@tempc \@tempb \let \@tempb \@tempa \fi \ifx
  \@tempb \@empty \def\@tempb {arXiv}\fi \@ifundefined
  {mn@eprint@\@tempb}{\@tempb:\@tempc}{\expandafter \expandafter \csname
  mn@eprint@\@tempb\endcsname \expandafter{\@tempc}}}

\bibitem[\protect\citeauthoryear{Ahn et~al.,}{Ahn et~al.}{2014}]{Ahn2014}
Ahn C.~P.,  et~al., 2014, \mn@doi [\apjs] {10/gf8n23}, 211, 17

\bibitem[\protect\citeauthoryear{Alexander}{Alexander}{2002}]{Alexander2002}
Alexander P.,  2002, \mn@doi [\mnras] {10/bcndwp}, 335, 610

\bibitem[\protect\citeauthoryear{Alexander}{Alexander}{2006}]{Alexander2006a}
Alexander P.,  2006, \mn@doi [\mnras] {10/fjrf8t}, 368, 1404

\bibitem[\protect\citeauthoryear{Alexander \& Hickox}{Alexander \&
  Hickox}{2012}]{Alexander2012}
Alexander D.,  Hickox R.,  2012, \mn@doi [New Astron. Rev.] {10/fwqx3f}, 56, 93

\bibitem[\protect\citeauthoryear{{Astropy Collaboration} et~al.,}{{Astropy
  Collaboration} et~al.}{2013}]{AstropyCollaboration2013}
{Astropy Collaboration} et~al., 2013, \mn@doi [A\&A] {10/gfvntd}, 558, A33

\bibitem[\protect\citeauthoryear{{Astropy Collaboration} et~al.,}{{Astropy
  Collaboration} et~al.}{2018}]{AstropyCollaboration2018}
{Astropy Collaboration} et~al., 2018, \mn@doi [\aj] {10/gfvntf}, 156, 123

\bibitem[\protect\citeauthoryear{Baldi, Capetti  \& Giovannini}{Baldi
  et~al.}{2019}]{Baldi2019a}
Baldi R.~D.,  Capetti A.,   Giovannini G.,  2019, \mn@doi [\mnras] {10/gh65v7},
  482, 2294

\bibitem[\protect\citeauthoryear{Banfield et~al.,}{Banfield
  et~al.}{2015}]{Banfield2015}
Banfield J.~K.,  et~al., 2015, \mn@doi [\mnras] {10/f7xbjh}, 453, 2326

\bibitem[\protect\citeauthoryear{Becker, White  \& Helfand}{Becker
  et~al.}{1995}]{Becker1995}
Becker R.~H.,  White R.~L.,   Helfand D.~J.,  1995, \mn@doi [\apj] {10/dr94br},
  450, 559

\bibitem[\protect\citeauthoryear{Begelman \& Cioffi}{Begelman \&
  Cioffi}{1989}]{Begelman1989}
Begelman M.~C.,  Cioffi D.~F.,  1989, \mn@doi [\apj] {10/d85x6k}, 345, L21

\bibitem[\protect\citeauthoryear{Bennett \& Simth}{Bennett \&
  Simth}{1962}]{Bennett1962}
Bennett A.~S.,  Simth F.~G.,  1962, \mn@doi [\mnras] {10/gfx7t4}, 125, 75

\bibitem[\protect\citeauthoryear{Bicknell}{Bicknell}{1994}]{Bicknell1994}
Bicknell G.~V.,  1994, \mn@doi [\apj] {10/fjfkwx}, 422, 542

\bibitem[\protect\citeauthoryear{Bicknell}{Bicknell}{1995}]{Bicknell1995}
Bicknell G.~V.,  1995, \mn@doi [\apjs] {10/bkwq6d}, 101, 29

\bibitem[\protect\citeauthoryear{Bicknell, Mukherjee, Wagner, Sutherland  \&
  Nesvadba}{Bicknell et~al.}{2018}]{Bicknell2018}
Bicknell G.~V.,  Mukherjee D.,  Wagner A.~Y.,  Sutherland R.~S.,   Nesvadba N.
  P.~H.,  2018, \mn@doi [\mnras] {10/gdb7mf}, 475, 3493

\bibitem[\protect\citeauthoryear{Bourne \& Sijacki}{Bourne \&
  Sijacki}{2017}]{Bourne2017}
Bourne M.~A.,  Sijacki D.,  2017, \mn@doi [\mnras] {10/gcnrtb}, 472, 4707

\bibitem[\protect\citeauthoryear{Bromberg, Nakar, Piran  \& Sari}{Bromberg
  et~al.}{2011}]{Bromberg2011}
Bromberg O.,  Nakar E.,  Piran T.,   Sari R.,  2011, \mn@doi [\apj]
  {10/c8272v}, 740, 100

\bibitem[\protect\citeauthoryear{Bruni et~al.,}{Bruni et~al.}{2020}]{Bruni2020}
Bruni G.,  et~al., 2020, \mn@doi [\mnras] {10/gh65v9}, 494, 902

\bibitem[\protect\citeauthoryear{Capetti, Baldi, Brienza, Morganti  \&
  Giovannini}{Capetti et~al.}{2019}]{Capetti2019}
Capetti A.,  Baldi R.~D.,  Brienza M.,  Morganti R.,   Giovannini G.,  2019,
  \mn@doi [\aap] {10/gh657t}, 631, A176

\bibitem[\protect\citeauthoryear{Cavaliere \& {Fusco-Femiano}}{Cavaliere \&
  {Fusco-Femiano}}{1978}]{Cavaliere1978}
Cavaliere A.,  {Fusco-Femiano} R.,  1978, \aap, 70, 677

\bibitem[\protect\citeauthoryear{Croton et~al.,}{Croton
  et~al.}{2006}]{Croton2006}
Croton D.~J.,  et~al., 2006, \mn@doi [\mnras] {10/dczprs}, 365, 11

\bibitem[\protect\citeauthoryear{Dubois, Devriendt, Slyz  \& Teyssier}{Dubois
  et~al.}{2012}]{Dubois2012}
Dubois Y.,  Devriendt J.,  Slyz A.,   Teyssier R.,  2012, \mn@doi [\mnras]
  {10/fzpb3w}, 420, 2662

\bibitem[\protect\citeauthoryear{English, Hardcastle  \& Krause}{English
  et~al.}{2016}]{English2016}
English W.,  Hardcastle M.~J.,   Krause M. G.~H.,  2016, \mn@doi [\mnras]
  {10/f84jr2}, 461, 2025

\bibitem[\protect\citeauthoryear{English, Hardcastle  \& Krause}{English
  et~al.}{2019}]{English2019}
English W.,  Hardcastle M.~J.,   Krause M. G.~H.,  2019, \mn@doi [\mnras]
  {10/gh657v}, 490, 5807

\bibitem[\protect\citeauthoryear{Fabian}{Fabian}{2012}]{Fabian2012}
Fabian A.,  2012, \mn@doi [\araa] {10/gfx7s9}, 50, 455

\bibitem[\protect\citeauthoryear{Falle}{Falle}{1991}]{Falle1991}
Falle S. A. E.~G.,  1991, \mn@doi [\mnras] {10/gfx7st}, 250, 581

\bibitem[\protect\citeauthoryear{Fanaroff \& Riley}{Fanaroff \&
  Riley}{1974}]{Fanaroff1974}
Fanaroff B.~L.,  Riley J.~M.,  1974, \mn@doi [\mnras] {10/gfx7vm}, 167, 31P

\bibitem[\protect\citeauthoryear{Gaibler, Krause  \& Camenzind}{Gaibler
  et~al.}{2009}]{Gaibler2009}
Gaibler V.,  Krause M.,   Camenzind M.,  2009, \mn@doi [\mnras] {10/d7wjkz},
  400, 1785

\bibitem[\protect\citeauthoryear{Gaibler, Khochfar  \& Krause}{Gaibler
  et~al.}{2011}]{Gaibler2011a}
Gaibler V.,  Khochfar S.,   Krause M.,  2011, \mn@doi [\mnras] {10/c82p79},
  411, 155

\bibitem[\protect\citeauthoryear{Gaibler, Khochfar, Krause  \& Silk}{Gaibler
  et~al.}{2012}]{Gaibler2012}
Gaibler V.,  Khochfar S.,  Krause M.,   Silk J.,  2012, \mn@doi [\mnras]
  {10/f373pc}, 425, 438

\bibitem[\protect\citeauthoryear{Garofalo \& Singh}{Garofalo \&
  Singh}{2019}]{Garofalo2019}
Garofalo D.,  Singh C.~B.,  2019, \mn@doi [\apj] {10/ghf3x7}, 871, 259

\bibitem[\protect\citeauthoryear{Hardcastle}{Hardcastle}{2018}]{Hardcastle2018}
Hardcastle M.~J.,  2018, \mn@doi [\mnras] {10/gc8k4b}, 475, 2768

\bibitem[\protect\citeauthoryear{Hardcastle \& Croston}{Hardcastle \&
  Croston}{2020}]{Hardcastle2020}
Hardcastle M.~J.,  Croston J.~H.,  2020, \mn@doi [\nar] {10/ghp9wx}, 88, 101539

\bibitem[\protect\citeauthoryear{Hardcastle \& Krause}{Hardcastle \&
  Krause}{2013}]{Hardcastle2013}
Hardcastle M.~J.,  Krause M. G.~H.,  2013, \mn@doi [\mnras] {10/f6dvz4}, 430,
  174

\bibitem[\protect\citeauthoryear{Hardcastle \& Krause}{Hardcastle \&
  Krause}{2014}]{Hardcastle2014}
Hardcastle M.~J.,  Krause M. G.~H.,  2014, \mn@doi [\mnras] {10/f6dvz4}, 443,
  1482

\bibitem[\protect\citeauthoryear{Hardcastle et~al.,}{Hardcastle
  et~al.}{2016}]{Hardcastle2016}
Hardcastle M.~J.,  et~al., 2016, \mn@doi [\mnras] {10/f77jmr}, 455, 3526

\bibitem[\protect\citeauthoryear{Hardcastle et~al.,}{Hardcastle
  et~al.}{2019a}]{Hardcastle2019a}
Hardcastle M.~J.,  et~al., 2019a, \mn@doi [\mnras] {10/ghhxd9}, 488, 3416

\bibitem[\protect\citeauthoryear{Hardcastle et~al.,}{Hardcastle
  et~al.}{2019b}]{Hardcastle2019}
Hardcastle M.~J.,  et~al., 2019b, \mn@doi [\aap] {10/ghf3x8}, 622, A12

\bibitem[\protect\citeauthoryear{Harris et~al.,}{Harris et~al.}{2020}]{NumPy}
Harris C.~R.,  et~al., 2020, \mn@doi [\nat] {10/ghbzf2}, 585, 357

\bibitem[\protect\citeauthoryear{Harwood, Vernstrom  \& Stroe}{Harwood
  et~al.}{2020}]{Harwood2020}
Harwood J.~J.,  Vernstrom T.,   Stroe A.,  2020, \mn@doi [\mnras] {10/ghbxmr},
  491, 803

\bibitem[\protect\citeauthoryear{Horton, Hardcastle, Read  \& Krause}{Horton
  et~al.}{2020a}]{Horton2020a}
Horton M.~A.,  Hardcastle M.~J.,  Read S.~C.,   Krause M. G.~H.,  2020a,
  \mn@doi [\mnras] {10/ghkhff}, 493, 3911

\bibitem[\protect\citeauthoryear{Horton, Krause  \& Hardcastle}{Horton
  et~al.}{2020b}]{Horton2020}
Horton M.~A.,  Krause M. G.~H.,   Hardcastle M.~J.,  2020b, \mn@doi [\mnras]
  {10/ghkhfd}, 499, 5765

\bibitem[\protect\citeauthoryear{Hunter}{Hunter}{2007}]{Matplotlib}
Hunter J.~D.,  2007, \mn@doi [Computing in Science \& Engineering] {10/drbjhg},
  9, 90

\bibitem[\protect\citeauthoryear{Joshi et~al.,}{Joshi et~al.}{2019}]{Joshi2019}
Joshi R.,  et~al., 2019, \mn@doi [\apj] {10/gg3ff7}, 887, 266

\bibitem[\protect\citeauthoryear{Kaiser \& Alexander}{Kaiser \&
  Alexander}{1997}]{Kaiser1997}
Kaiser C.~R.,  Alexander P.,  1997, \mn@doi [\mnras] {10/gfx7zd}, 286, 215

\bibitem[\protect\citeauthoryear{Kaiser, {Dennett-Thorpe}  \& Alexander}{Kaiser
  et~al.}{1997}]{Kaiser1997a}
Kaiser C.~R.,  {Dennett-Thorpe} J.,   Alexander P.,  1997, \mn@doi [\mnras]
  {10/gfx7zf}, 292, 723

\bibitem[\protect\citeauthoryear{Kluyver et~al.,}{Kluyver
  et~al.}{2016}]{Jupyter}
Kluyver T.,  et~al., 2016, in Loizides F.,  Scmidt B.,  eds, Positioning and
  Power in Academic Publishing: Players, Agents and Agendas. {IOS Press},
  {Netherlands}, pp 87--90

\bibitem[\protect\citeauthoryear{Krause}{Krause}{2003}]{Krause2003}
Krause M.,  2003, \mn@doi [\aap] {10/bm39r5}, 398, 113

\bibitem[\protect\citeauthoryear{Krause}{Krause}{2005}]{Krause2005}
Krause M.,  2005, \mn@doi [A\&A] {10/bczq7k}, 431, 45

\bibitem[\protect\citeauthoryear{Krause, Alexander, Riley  \& Hopton}{Krause
  et~al.}{2012}]{Krause2012}
Krause M.,  Alexander P.,  Riley J.,   Hopton D.,  2012, \mn@doi [\mnras]
  {10/f4m6zm}, 427, 3196

\bibitem[\protect\citeauthoryear{Krause et~al.,}{Krause
  et~al.}{2019}]{Krause2019a}
Krause M. G.~H.,  et~al., 2019, \mn@doi [\mnras] {10/gfx7vx}, 482, 240

\bibitem[\protect\citeauthoryear{Li, Wiita, Schuh, Elghossain  \& Hu}{Li
  et~al.}{2018}]{Li2018a}
Li Y.,  Wiita P.~J.,  Schuh T.,  Elghossain G.,   Hu S.,  2018, \mn@doi [\apj]
  {10/ggckp5}, 869, 32

\bibitem[\protect\citeauthoryear{Longair}{Longair}{2011}]{Longair2011}
Longair M.~S.,  2011, High {{Energy Astrophysics}}, 3rd edn.
{Cambridge University Press}

\bibitem[\protect\citeauthoryear{Mahatma et~al.,}{Mahatma
  et~al.}{2018}]{Mahatma2018}
Mahatma V.~H.,  et~al., 2018, \mn@doi [\mnras] {10/gdb4pv}, 475, 4557

\bibitem[\protect\citeauthoryear{Mart{\'i}, M{\"u}ller, Font, Ib{\'a}{\~n}ez
  \& Marquina}{Mart{\'i} et~al.}{1997}]{Marti1997}
Mart{\'i} J.~M.,  M{\"u}ller E.,  Font J.~A.,  Ib{\'a}{\~n}ez J. M.~Z.,
  Marquina A.,  1997, \mn@doi [\apj] {10/b5332w}, 479, 151

\bibitem[\protect\citeauthoryear{Massaglia, Bodo, Rossi, Capetti  \&
  Mignone}{Massaglia et~al.}{2016}]{Massaglia2016}
Massaglia S.,  Bodo G.,  Rossi P.,  Capetti S.,   Mignone A.,  2016, \mn@doi
  [\aap] {10/gfx7xk}, 12, 1

\bibitem[\protect\citeauthoryear{Mathews}{Mathews}{1971}]{Mathews1971}
Mathews W.~G.,  1971, \mn@doi [\apj] {10/fhv22r}, 165, 147

\bibitem[\protect\citeauthoryear{Matsumoto, Komissarov  \&
  Gourgouliatos}{Matsumoto et~al.}{2021}]{Matsumoto2021}
Matsumoto J.,  Komissarov S.~S.,   Gourgouliatos K.~N.,  2021, \mn@doi [\mnras]
  {10/gk7q7f}, 503, 4918

\bibitem[\protect\citeauthoryear{Matthews, Bell  \& Blundell}{Matthews
  et~al.}{2020}]{Matthews2020}
Matthews J.~H.,  Bell A.~R.,   Blundell K.~M.,  2020, \mn@doi [\nar]
  {10/gh66k8}, 89, 101543

\bibitem[\protect\citeauthoryear{McNamara \& Nulsen}{McNamara \&
  Nulsen}{2007}]{McNamara2007}
McNamara B.,  Nulsen P.,  2007, \mn@doi [\araa] {10/bhm45r}, 45, 117

\bibitem[\protect\citeauthoryear{Mendygral, Jones  \& Dolag}{Mendygral
  et~al.}{2012}]{Mendygral2012}
Mendygral P.~J.,  Jones T.~W.,   Dolag K.,  2012, \mn@doi [\apj] {10/ggk8rk},
  750, 166

\bibitem[\protect\citeauthoryear{Middelberg et~al.,}{Middelberg
  et~al.}{2008}]{Middelberg2008}
Middelberg E.,  et~al., 2008, \mn@doi [\aj] {10/fvckk4}, 135, 1276

\bibitem[\protect\citeauthoryear{Mignone \& McKinney}{Mignone \&
  McKinney}{2007}]{Mignone2007a}
Mignone A.,  McKinney J.~C.,  2007, \mn@doi [\mnras] {10/cppcfb}, 378, 1118

\bibitem[\protect\citeauthoryear{Mignone, Bodo, Massaglia, Matsakos, Tesileanu,
  Zanni  \& Ferrari}{Mignone et~al.}{2007}]{Mignone2007}
Mignone A.,  Bodo G.,  Massaglia S.,  Matsakos T.,  Tesileanu O.,  Zanni C.,
  Ferrari A.,  2007, \mn@doi [\apjs] {10/fhh3hj}, 170, 228

\bibitem[\protect\citeauthoryear{Mingo et~al.,}{Mingo et~al.}{2019}]{Mingo2019}
Mingo B.,  et~al., 2019, \mn@doi [\mnras] {10/gf66tb}, 488, 2701

\bibitem[\protect\citeauthoryear{Mukherjee, Bicknell, Sutherland  \&
  Wagner}{Mukherjee et~al.}{2016}]{Mukherjee2016}
Mukherjee D.,  Bicknell G.~V.,  Sutherland R.,   Wagner A.,  2016, \mn@doi
  [\mnras] {10/f8489c}, 461, 967

\bibitem[\protect\citeauthoryear{Mukherjee, Bodo, Mignone, Rossi  \&
  Vaidya}{Mukherjee et~al.}{2020}]{Mukherjee2020}
Mukherjee D.,  Bodo G.,  Mignone A.,  Rossi P.,   Vaidya B.,  2020, \mn@doi
  [\mnras] {10/ghkhfx}, 499, 681

\bibitem[\protect\citeauthoryear{Mullin, Riley  \& Hardcastle}{Mullin
  et~al.}{2008}]{Mullin2008}
Mullin L.~M.,  Riley J.~M.,   Hardcastle M.~J.,  2008, \mn@doi [\mnras]
  {10/cx94t6}, 390, 595

\bibitem[\protect\citeauthoryear{Norris et~al.,}{Norris
  et~al.}{2006}]{Norris2006}
Norris R.~P.,  et~al., 2006, \mn@doi [\aj] {10/dbtsvd}, 132, 2409

\bibitem[\protect\citeauthoryear{O'Neill, Jones, Nolting  \& Mendygral}{O'Neill
  et~al.}{2019}]{ONeill2019}
O'Neill B.~J.,  Jones T.~W.,  Nolting C.,   Mendygral P.~J.,  2019, \mn@doi
  [\apj] {10/ghhxg7}, 884, 12

\bibitem[\protect\citeauthoryear{Perucho, Mart{\'i}  \& Quilis}{Perucho
  et~al.}{2019}]{Perucho2019b}
Perucho M.,  Mart{\'i} J.-M.,   Quilis V.,  2019, \mn@doi [\mnras] {10/gfx7ss},
  482, 3718

\bibitem[\protect\citeauthoryear{{Planck Collaboration} et~al.,}{{Planck
  Collaboration} et~al.}{2016}]{Planck2016}
{Planck Collaboration} et~al., 2016, \mn@doi [A\&A] {10/f9scmm}, 594, A13

\bibitem[\protect\citeauthoryear{Rafferty, McNamara, Nulsen  \& Wise}{Rafferty
  et~al.}{2006}]{Rafferty2006}
Rafferty D.~A.,  McNamara B.~R.,  Nulsen P. E.~J.,   Wise M.~W.,  2006, \mn@doi
  [\apj] {10/c3jdct}, 652, 216

\bibitem[\protect\citeauthoryear{Rafferty, McNamara  \& Nulsen}{Rafferty
  et~al.}{2008}]{Rafferty2008}
Rafferty D.~A.,  McNamara B.~R.,   Nulsen P. E.~J.,  2008, \mn@doi [\apj]
  {10/bwdn6j}, 687, 899

\bibitem[\protect\citeauthoryear{Raouf, Shabala, Croton, Khosroshahi  \&
  Bernyk}{Raouf et~al.}{2017}]{Raouf2017}
Raouf M.,  Shabala S.~S.,  Croton D.~J.,  Khosroshahi H.~G.,   Bernyk M.,
  2017, \mn@doi [\mnras] {10/gbvpjc}, 471, 658

\bibitem[\protect\citeauthoryear{Raouf, Silk, Shabala, Mamon, Croton,
  Khosroshahi  \& Beckmann}{Raouf et~al.}{2019}]{Raouf2019}
Raouf M.,  Silk J.,  Shabala S.~S.,  Mamon G.~A.,  Croton D.~J.,  Khosroshahi
  H.~G.,   Beckmann R.~S.,  2019, \mn@doi [\mnras] {10/gfzv5m}, 486, 1509

\bibitem[\protect\citeauthoryear{Rieger \& Duffy}{Rieger \&
  Duffy}{2021}]{Rieger2021}
Rieger F.~M.,  Duffy P.,  2021, \mn@doi [\apj] {10/gh24ph}, 907, L2

\bibitem[\protect\citeauthoryear{Rodman et~al.,}{Rodman
  et~al.}{2019}]{Rodman2019}
Rodman P.~E.,  et~al., 2019, \mn@doi [\mnras] {10/gfx7xd}, 482, 5625

\bibitem[\protect\citeauthoryear{Scheuer}{Scheuer}{1974}]{Scheuer1974}
Scheuer P. a.~G.,  1974, \mn@doi [\mnras] {10/gfx7v6}, 166, 513

\bibitem[\protect\citeauthoryear{Schindler, Binggeli  \&
  B{\"o}hringer}{Schindler et~al.}{1999}]{Schindler1999}
Schindler S.,  Binggeli B.,   B{\"o}hringer H.,  1999, \aap, 343, 420

\bibitem[\protect\citeauthoryear{Seymour et~al.,}{Seymour
  et~al.}{2020}]{Seymour2020}
Seymour N.,  et~al., 2020, \mn@doi [\pasa] {10/ggq34w}, 37, e013

\bibitem[\protect\citeauthoryear{Shabala \& Alexander}{Shabala \&
  Alexander}{2009}]{Shabala2009}
Shabala S.,  Alexander P.,  2009, \mn@doi [\apj] {10/cnhv3h}, 699, 525

\bibitem[\protect\citeauthoryear{Shabala \& Godfrey}{Shabala \&
  Godfrey}{2013}]{Shabala2013a}
Shabala S.~S.,  Godfrey L. E.~H.,  2013, \mn@doi [\apj] {10/gfx7w2}, 769, 129

\bibitem[\protect\citeauthoryear{Shabala, Jurlin, Morganti, Brienza,
  Hardcastle, Godfrey, Krause  \& Turner}{Shabala et~al.}{2020}]{Shabala2020}
Shabala S.~S.,  Jurlin N.,  Morganti R.,  Brienza M.,  Hardcastle M.~J.,
  Godfrey L. E.~H.,  Krause M. G.~H.,   Turner R.~J.,  2020, \mn@doi [\mnras]
  {10/ggt8qh}

\bibitem[\protect\citeauthoryear{Sutherland \& Dopita}{Sutherland \&
  Dopita}{1993}]{Sutherland1993}
Sutherland R.~S.,  Dopita M.~A.,  1993, \mn@doi [\apjs] {10/cwm2cc}, 88, 253

\bibitem[\protect\citeauthoryear{Taub}{Taub}{1948}]{Taub1948}
Taub A.~H.,  1948, \mn@doi [Phys. Rev.] {10/cx8v32}, 74, 328

\bibitem[\protect\citeauthoryear{Turner \& Shabala}{Turner \&
  Shabala}{2015}]{Turner2015}
Turner R.~J.,  Shabala S.~S.,  2015, \mn@doi [\apj] {10/gfx7vg}, 806, 59

\bibitem[\protect\citeauthoryear{Turner, Rogers, Shabala  \& Krause}{Turner
  et~al.}{2018a}]{Turner2018a}
Turner R.~J.,  Rogers J.~G.,  Shabala S.~S.,   Krause M. G.~H.,  2018a, \mn@doi
  [\mnras] {10/gcz46k}, 473, 4179

\bibitem[\protect\citeauthoryear{Turner, Shabala  \& Krause}{Turner
  et~al.}{2018b}]{Turner2018}
Turner R.~J.,  Shabala S.~S.,   Krause M. G.~H.,  2018b, \mn@doi [\mnras]
  {10/gc2482}, 474, 3361

\bibitem[\protect\citeauthoryear{Vikhlinin, Kravtsov, Forman, Jones,
  Markevitch, Murray  \& Van~Speybroeck}{Vikhlinin
  et~al.}{2006}]{Vikhlinin2006}
Vikhlinin A.,  Kravtsov A.,  Forman W.,  Jones C.,  Markevitch M.,  Murray
  S.~S.,   Van~Speybroeck L.,  2006, \mn@doi [\apj] {10/dvbjh7}, 640, 691

\bibitem[\protect\citeauthoryear{Virtanen et~al.,}{Virtanen
  et~al.}{2020}]{SciPy}
Virtanen P.,  et~al., 2020, \mn@doi [Nature Methods] {10/ggj45f}, 17, 261

\bibitem[\protect\citeauthoryear{Wagner, Bicknell  \& Umemura}{Wagner
  et~al.}{2012}]{Wagner2012a}
Wagner A.~Y.,  Bicknell G.~V.,   Umemura M.,  2012, \mn@doi [\apj] {10/gfx7tf},
  757, 136

\bibitem[\protect\citeauthoryear{Weinberger et~al.,}{Weinberger
  et~al.}{2018}]{Weinberger2017}
Weinberger R.,  et~al., 2018, \mn@doi [\mnras] {10/gd7b8w}, 479, 4056

\bibitem[\protect\citeauthoryear{Yates, Shabala  \& Krause}{Yates
  et~al.}{2018}]{Yates2018}
Yates P.~M.,  Shabala S.~S.,   Krause M. G.~H.,  2018, \mn@doi [\mnras]
  {10/gfn2h4}, 480, 5286

\makeatother
\end{thebibliography}

  %%%%%%%%%%%%%%%%%%%%%%%%%%%%%%%%%%%%%%%%%%%%%%%%%%

  %%%%%%%%%%%%%%%%% APPENDICES %%%%%%%%%%%%%%%%%%%%%

  %%%%%%%%%%%%%%%%%%%%%%%%%%%%%%%%%%%%%%%%%%%%%%%%%%

  % Don't change these lines
  \bsp	% typesetting comment
  \label{lastpage}
\end{document}